\documentclass[journal=jctc,manuscript=article]{achemso}

\usepackage[version=3]{mhchem}
\usepackage{multirow}
\usepackage{colortbl}
\usepackage{xcolor}
\usepackage{booktabs}
\usepackage{nicefrac}

\definecolor{Prussian}{HTML}{05026F}
\definecolor{Salmon}{HTML}{F98458}
\definecolor{Sea}{HTML}{AACCBF}
\definecolor{LightGrey}{HTML}{EDEDED}
\definecolor{Barbie}{HTML}{A82C5C}
\definecolor{OldRose}{HTML}{CDAEA4}
\definecolor{comment}{HTML}{4D31D8}
\definecolor{QEred}{HTML}{B11815}

\newcommand*{\citen}[1]{\begingroup \romannumeral-`\x \setcitestyle{numbers}\cite{#1}\endgroup}
\newcommand*{\celco}[1]{\cellcolor[HTML]{#1}}
\newcommand*{\tcw}{\color{white}}
\newcommand*{\QE}{\textsc{Quantum} ESPRESSO}
\newcommand{\overhang}{\dimexpr\tabcolsep+0.1pt\relax}
\renewcommand*{\arraystretch}{1.2}
\newcommand{\U}{\textrm{U}}
\newcommand{\J}{\textrm{J}}
\newcommand*{\DeltaE}{$\Delta E_\textrm{HL}$}
\newcommand{\etal}{\textit{et al}.}
\newcommand{\insitu}{\textit{in situ}}
\newcommand{\videinfra}{\textit{vide infra}}
\let\tr\textrm

\usepackage{listings}
\lstdefinestyle{mystyle}{
    backgroundcolor=\color{LightGrey},   
    commentstyle=\color{comment},
    keywordstyle=\color{blue},
    stringstyle=\color{QEred},
    showstringspaces=false,
    basicstyle=\ttfamily\tiny}
\author{Lórien MacEnulty}
\affiliation[TCD]
{School of Physics, CRANN Institute, 
and  AMBER Centre, Trinity College Dublin, The University of Dublin, Dublin 2, Ireland}
\email{lmacenul@tcd.ie}

\author{João Paulo Almeida de Mendonça}
\affiliation[UGA]
{Université Grenoble Alpes, CNRS, Grenoble INP, SIMaP, 38000 Grenoble, France}
\author{Roberta Poloni}
\email{roberta.poloni@grenoble-inp.fr}
\author{David D. O'Regan}
\affiliation[TCD]
{School of Physics, CRANN Institute, 
and  AMBER Centre, Trinity College Dublin, The University of Dublin, Dublin 2, Ireland}

\title[Benchmarking total energies with Hund's J terms in Hubbard-corrected spin-crossover chemistry]
  {Benchmarking total energies with Hund's J terms in Hubbard-corrected spin-crossover chemistry}


\begin{document}

\begin{tocentry}
\centering
\includegraphics[trim={0.25cm 0.25cm 0.05cm 0.3cm},clip,width=1.0\linewidth]{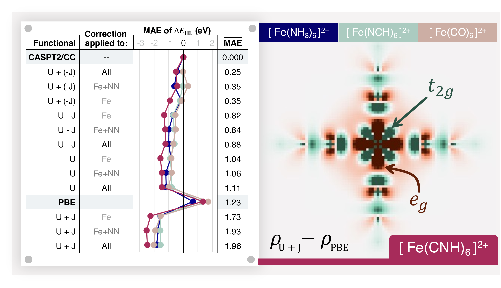} 

\end{tocentry}

\begin{abstract}

The effect of the Hund's J terms in various DFT+U+J corrections to semi-local spin-density functional theory is assessed for a series of four octahedrally-coordinated Fe(II) spin-crossover molecules spanning the covalent end of the ligand field spectrum. We report values and analyze trends for the Hubbard U and Hund’s J parameters determined via minimum-tracking linear response for all valence atomic subspaces and relevant spin states in these molecules. We then methodically apply them via simplified rotationally-invariant Hubbard functionals in search of the simplest combination to yield reliable adiabatic energy differences with respect to those obtained using CASPT2/CC. The observed failure of canonical, positively-signed Hund’s J terms in furthering the already robust capacity
of DFT+U to obtain accurate energetics prompts an evaluation of their limitations when seeking to account for the  static correlation phenomena in such strongly covalent systems and suggests directions for their improvement.

\end{abstract}

\section{Introduction}

An emerging class of technology based on spin crossover (SCO)—a quantum mechanism of material systems that can engender dramatic physical and chemical changes—exhibits high potential for eco-friendly and renewable applications like carbon capture\cite{Culp2013,MarianoThesis,Mohajeri2020,Mariano2023,KARADAS2012,ogilvie2013}, energy storage\cite{wang_prussian_2018}, spintronics\cite{FENG2019,magnetochemistry2010018,Kipgen2021} and sensor devices.\cite{Cheng2023,thallapally_prussian_2010,matos2019} The material systems that demonstrate SCO---from mononuclear molecules\cite{Murray2013} and metal-organic frameworks \cite{Pham2016,Gutlich2004,Costa2025} to bulk solids\cite{Pinkowicz2015} like Prussian Blue analogues\cite{wang_prussian_2018,Kabir2012,KARADAS2012,Middlemiss2008,Papanikolaou2006,Aguila2016,matos2019}---may reversibly switch states from the low spin (LS) to their high spin (HS) configurations. This switch may be stimulated by various environmental changes, including the presence of guest molecules\cite{NI201728,Gutlich2004,Murray2013}, magnetic and electric fields\cite{sato2004}, temperature\cite{Aguila2016,GUTLICH19901,Warner2013}, pressure\cite{papanikolaou2007,Pinkowicz2015}, and light irradiation.\cite{Cheng2023,Papanikolaou2006,GUTLICH19901,Warner2013}

A quantity of interest to those looking to harness SCO is the critical temperature $T_{1/2}$, the dominant contribution of which is proportional to the adiabatic energy difference \DeltaE$= E_\textrm{HS}-E_\textrm{LS}$.\cite{Vidal2021} Without an accurate description of the electronic structure of SCO materials and its influence on \DeltaE, it is difficult to design and discover new SCO materials that optimize and tailor functionality while simultaneously minimizing, if not eliminating, impracticalities and impediments.

Materials scientists and quantum chemists often look to density functional theory (DFT) for this task\cite{Swart2008,Cirera2012,Vidal2021,Middlemiss2008,Pierloot2006}, especially as \DeltaE\ falls increasingly within approximate DFT’s regime of achievable accuracy. Varying percentages of Hartree-Fock (HF) exact exchange mixed with (semi-)local exchange-correlation functionals in DFT\cite{song2018} have been shown to obtain adiabatic energy differences in good agreement with the more expensive wavefunction-based methods\cite{radon2019,Domingo2010} like quantum Monte Carlo\cite{KARADAS2012,droghetti_assessment_2012,song2018,Fumanal2016}, CASPT2\cite{Pierloot2006,Pierloot2017,Mariano2021,Domingo2010} and/or coupled cluster\cite{Mariano2021}. Furthermore, as some of the present authors have recently shown, artificial neural networks can be leveraged to develop exchange and correlation functionals that yield values of adiabatic energy differences comparable to highly accurate quantum chemistry methods.\cite{Joao2023} Yet, many of the aforementioned methods are either computationally expensive, non-generalizable, or otherwise inaccessible for routine or high-throughput use. 

Parallel works among the wider materials electronic structure theory community have highlighted (first-principles parameters) DFT+U(+J)\cite{Mariano2020,Mariano2021,Mariano2023,MarianoThesis,hegner2016,wojdel2008,wojdel2009,Fabris2013} as a practical alternative having demonstrated restoration of electronic structures\cite{anisimov_band_1991,anisimov_density_1991,anisimov_density_1993,orhan_first-principles_2020,Lambert2023,georges_strong_2013} and total energy differences\cite{Wang2021,Yu2020,Dorado2009,Dorado2010,Tompsett2012,Patrick2016,Gopal2017,MacEnulty2023,burgess2023} of Mott-Hubbard systems without increasing computation time 
significantly. 

Despite the longstanding competence of DFT+U and the linear response approach to calculating its parameters \emph{in situ}, these methods still foster many unanswered questions, especially in the context of spin-crossover chemistry.
These are related, in particular, to the comparability of DFT+U total energies for non-ground state magnetic systems. Earlier work has shown that in the case of Fe(II) complexes, DFT+U with Hubbard corrections applied to Fe atoms yield adiabatic energy differences that may deviate, by several eV,\cite{Mariano2020,Mariano2023} from quantum chemistry methods. The origin of this issue is attributed to a substantially larger Hubbard energy correction applied to the LS state, thus biasing the calculations towards the HS state and yielding too-negative values of \DeltaE. A follow-up study by the same authors demonstrated that such bias can be overcome by adopting a \emph{density-corrected scheme} (otherwise known as PBE@$f$, where $f$ is a Hubbard functional), where the Hubbard energy terms are used to compute the density before they are removed non-self-consistently from the total energies.\cite{Mariano2021}.

The use of a non-self-consistent scheme is one of several equally important questions related to fully self-consistent DFT+U calculations. For example, the first-principles Hubbard parameters have been shown to depend on the spin-state of the system\cite{MacEnultyThesis,MacEnulty2023,hegner2016,Mariano2020}, but are the resulting energy functionals comparable to one another through total energy differences? Moreover, it remains unclear precisely which combination of valence subspaces requires correction for optimized energetics. And crucially, do the Hund’s J energy terms—as mitigators of the U correction or as terms conventionally penalizing parallel spin alignment in their own right—help or hinder the accuracy of total energy differences in practice?

\begin{figure}[t!]
    \centering
    \includegraphics[trim={0cm 0cm 0.3cm 0cm},clip,width=1.0\linewidth]{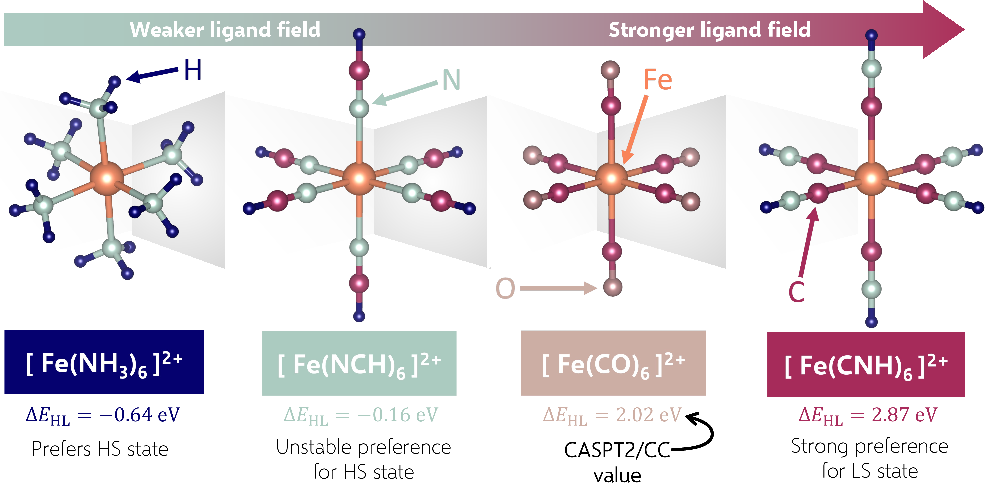} 
    \caption{Illustration of the four octahedrally coordinated \textrm{Fe(II)} complexes studies in the work. The reported adiabatic energy differences are from CASPT2/CC calculations (see text). Molecule images were generated with VESTA\cite{VESTA}.}
    \label{MoleculeGeoms}
\end{figure}

Given this context, we execute a study building on prior works\cite{Mariano2020,Mariano2021} designed to answer these and other questions. We first calculate, via minimum-tracking linear response,\cite{Linscott2018} and analyze trends of the Hubbard U and Hund’s J—to which we collectively refer as the Hubbard parameters (HPs)—for all atomic valence subspaces in a series of octahedrally-coordinated Fe(II) SCO molecules that span the more covalent end of the ligand-field strength spectrum (shown in Figure \ref{MoleculeGeoms}). Then, we methodically apply these parameters via the widely-used simplified rotationally invariant Hubbard functionals\cite{dudarev1998,himmetoglu2014} in search of the simplest combination to yield reliable adiabatic energy differences with respect to those obtained by a robust benchmark: the set of coupled cluster-corrected CASPT2 $\Delta E_\tr{HL}$ values from Ref.~\citen{Mariano2021}, which follow Refs.~\citen{Pierloot2017,Pierloot2018,Radon2023,Radon2024} in successfully employing the method to remove a well-known bias of CASPT2 towards the HS states.

A time-tested rotationally invariant formulation of DFT+U is known as the Dudarev functional,\cite{dudarev1998,Anisimov_1997} given by
\begin{equation}\label{sriform}
E_{\U-\J}[\{n_{mm'}^{i\sigma}\}]=\sum_{i\sigma} \frac{\U^i_\textrm{eff}}{2}\ \textrm{Tr}[\textbf{n}^{i\sigma} (1-\textbf{n}^{i\sigma})]\ ,
\end{equation}
where $\textbf{n}^{i\sigma}$ is the occupancy matrix (and $\{n_{mm'}^{i\sigma}\}$ its matrix elements) of the subspace $i=\{a,n,\ell \}$ (corresponding to fixed, pre-selected orbitals on atom $a$ with quantum numbers $n$ and $\ell$) undergoing correction. This formulation of DFT+U comprises a single corrective term per spin channel, which is added to the total energy, usually from a (semi-)local density functional approximation, and which penalizes fractional subspace  occupancy matrix eigenvalues for positive $\U^i_\tr{eff}$ values. The effective parameter is U$_\tr{eff}=\U-\J$, where U is the subspace-averaged density-density self-interaction, and J is the subspace-averaged exchange self-interaction, in practice calculated as a subspace-averaged spin-spin self-interaction. The Hund's J is included in Eq. \ref{sriform} as a mitigating coefficient weakening the strength of intra-site Coulomb correction to account for the effect of Hund's rules on spin and orbital polarization.\cite{Solovyev1994,georges_strong_2013} It is argued that the neglect of the Hund's J can result in excessive correction. In the literature, U$_{\tr{eff}}$ and U are often referenced interchangeably depending on the consideration given to J\cite{Kulik2015}, but it should be emphasized that the 
linear-response calculated U is not already U$_\tr{eff}$.\cite{Linscott2018}

An extension of the DFT+U formalism, derived by Himmetoglu \etal\cite{Himmetoglu2011}, includes the Hund's J also as a distinct correction aimed to better account for the correlation effect known as spin-flip exchange. In this form of DFT+U+J, which we call the Himmetoglu functional, the total correction is found to be (neglecting an optional minority-spin-specific term per established practice\cite{Himmetoglu2011,Orhan2020}) 
\begin{equation}\label{HimmetogluDFT+U+J}
E_{\U+\J}[\{n_{mm'}^{i\sigma}\}]=\sum_{i\sigma}{\frac{\U^i-\J^i}{2} \textrm{Tr} \left[\textbf{n}^{i\sigma}\left(\textbf{1}-\textbf{n}^{i\sigma}\right)\right]}+\sum_{i\sigma}{\frac{\J^i}{2}  \textrm{Tr} \left[\textbf{n}^{i\sigma}\textbf{n}^{i\bar{\sigma}}\right] }\ ,
\end{equation}
where $\bar{\sigma}$ is the opposite spin of spin $\sigma$. Here, we see that the Hund's J functions in two ways; (i) it mitigates the effect of U on the interactions between electrons with parallel spin, and (ii) it adds an explicit penalty for occupation of anti-aligned spins on the same spatial projector orbitals (more correctly, on the eigenstates of the opposite-spins' occupancy matrix product). The DFT+U+J potential operator acting on Kohn–Sham spin-${\sigma}$ states reads as

\begin{equation}\label{Himmetoglupot}
\hat{V}_{\U+\J}^\sigma = \sum_{m m'} \left[ (\U - \J) \left( \frac{1}{2} \delta_{mm'} - n_{mm'}^{\sigma} \right) + \J \, n_{mm'}^{\bar{\sigma}} \right] |\phi_m\rangle \langle \phi_{m'}|.
\end{equation}

Meanwhile, a very recently introduced family of Hubbard functionals called BLOR\cite{burgess2023,burgess2024} 
was derived directly from the flat-plane condition (not the Hubbard model), with therefore no need for a double-counting correction. While the direct use of BLOR is beyond the scope of our study as it  differs in several respects with respect to DFT+U, we have found some of the learning from that informative for our study. Specifically, in BLOR, the term scaled by J addresses static-correlation error (SCE) and turns out (unconventionally) to be negative in sign for positive J, acting to suppress local moments. In this context, it has been shown\cite{burgess2023,burgess2024} using dissociated molecular benchmark systems, that (positive for positive) J terms similar 
in form to item (ii) of Eq. \ref{HimmetogluDFT+U+J} have a sign that tends to push the energy away from its exact value in molecular systems. This is because these terms enhance localized spin moments by making the broken-symmetry spin-polarized system relatively more energetically favorable, but in doing so increase the total energy (and, in effect, reduce correlation). Spurious symmetry breaking may be acceptable if seeking to predict direct spectroscopic observables related to spin on single molecules, where the total energy is not as relevant, and it would not be problematic at all in solid-state systems exhibiting true spontaneous symmetry breaking in the thermodynamic limit. However, for properties like SCO that implicate ensembles of molecules, and certainly for comparing to quantum chemistry benchmarks without spurious symmetry breaking, it seems reasonable to consider the insights from BLOR. A question arises, in particular: if one were to simply (with no more justification than the above motivation from the BLOR functional) change the sign of J when employing the Himmetoglu DFT+U+J functional, would a similar effect occur and a more correct energy be recovered? We thus compare our results to a modified version of the Himmetoglu functional, denoted for concision in this work by DFT+U+(-J), in which we change the sign of the J parameter (which affects also 
the same-spin term). This experiment is intended as a potential proof of principle and, of course, not as a proposal for any wider adoption.

We test the following variants of Hubbard functionals: (i) DFT+U, (ii) DFT+U$_\tr{eff}$ (i.e.,~the Dudarev DFT+U-J), (iii) DFT+U+J (Himmetoglu), and (iv) DFT+U+(-J), with Hubbard parameters applied either to the iron atom alone, to iron and its nearest-neighboring shell, or to all atoms of the molecule. For each set, we employ two choices of Hubbard parameters: those determined \emph{in situ} for each spin state, and HS parameters for both LS and HS calculations.

We ultimately find that the simplest combination of techniques to yield reliable spin-state energetic properties, with respect to those obtained by CASPT2/CC, includes the Dudarev DFT+U$_\tr{eff}$ functional, applied only to the central iron atom, and using the same HP values regardless of the molecule's spin state. Our results illuminate the failure of the Hund’s J in furthering DFT+U’s already robust capacity to obtain accurate adiabatic energy differences. We thus map previously uncharted limitations of first-principles DFT+U+J and precisely highlight areas for improvement therein.

\section{\label{Sec:CompDetails}Computational Details}

We conduct our investigation on a series of four octahedrally coordinated Fe(II) complexes that span the covalent end of the spectrum of ligand field strengths: [Fe(NH$_3$)$_6$]$^{2+}$ (weakest ligand field), [Fe(NCH)$_6$]$^{2+}$, [Fe(CO)$_6$]$^{2+}$ and [Fe(CNH)$_6$]$^{2+}$ (strongest ligand field). The geometries of these molecules are optimized using the TPSSh functional\cite{TPSS,TPSSH2003} and are provided by  Ref.~\citen{Mariano2020}. We note that the spin-flip reorganization energy for these molecules is quite large, comparable to if not greater than their corresponding adiabatic energy differences.

As a benchmark method for calculating adiabatic energy differences, we use the coupled cluster-corrected CASPT2 (CASPT2/CC) \DeltaE\ values from Ref.~\citen{Mariano2021}. It is argued, there and in Refs.~\citen{Pierloot2017,Pierloot2018,Radon2023,Radon2024}, that this approach exploits CCSD(T) to improve the description of electronic correlation in the semi-core $3s3p$ states, which neutralizes CASPT2's tendency to overstabilize the HS over the LS states.

All calculations, linear response and Hubbard functionals, are spin polarized, and the total charge of the system is set to +2. We use the Perdew-Burke-Ernzerhof (PBE) GGA\cite{PBE1996} functional and the projector augmented wave (PAW) method\cite{blochl_projector_1994}, specifically the Jollet-Torrent-Holzwarth PBE PAW datasets (Version 1.0)\cite{Jollet2014} generated via \textit{atompaw}.\cite{holzwarth_projector_2001} Derived from those pseudopotentials, the PAW augmentation sphere cutoff radius $r_c$ used is 1.51 $a_0$ for C $2p$, 2.01 $a_0$ for Fe $3d$, 0.99 $a_0$ for H $1s$, 1.20 $a_0$ for N $2p$, and 1.41 $a_0$ for O $2p$.

Hubbard functional calculations are performed with the Order N Electronic Total Energy Package (ONETEP)\cite{ONETEP} using PBE and the PAW method\cite{blochl_projector_1994} to describe the exchange-correlation functional. Each complex is placed at the center of a cubic vacuum of side-length 75.59 $a_0$ (40 \AA). For ease of comparison with prior investigations,\cite{Mariano2020} a psinc basis set (see Refs.~\citen{Mostofi2002,Skylaris2001,Skylaris2002,psincbasis}) is selected to resemble a plane-wave basis set with a cutoff kinetic energy of 40 Ha and a fine-grid energy cutoff of 160 Ha. We facilitate convergence by enabling ensemble DFT with 0K smearing. The central Fe atom is described with a total of 26 non-orthogonal generalized Wannier functions (NGWFs) limited to a radius of 14 $a_0$; all $p$-block elements (i.e., N, C, O) are allotted eight NGWFs limited to a radius of 12 $a_0$, and H receives two NGWFs limited to 10 $a_0$. The NGWFs were initialized in split-norm pairs (see Refs.~\citen{psincbasis} and \citen{Artacho1999}), not least in order to afford more variational freedom in such spin-polarized systems. Total energies are converged to within $10^{-6}$ Ha (2.72$\times 10^{-5}$ eV). To correct for spurious electrostatic interactions between periodic images of the molecules, we use the Martyna-Tuckerman \cite{MartynaTuckerman} minimum image convention with cutoff of $7.0\ a_0$ following the suggestions of Hine \etal\cite{Hine2011}

For the linear response determination of the Hubbard U and Hund's J in ONETEP (discussed in more detail in Appendix A1 
and Ref.~\citen{MacEnultyThesis}), runtime parameters are the same as those used for functional calculations described above, except the cell size is set to 37.79 $a_0$. Four evenly-spaced linear response perturbations (following the minimum-tracking method of Ref.~\citen{Linscott2018}) ranging from -0.10 to 0.10 eV were applied. The zero-strength perturbation is also considered in the regression. In order to include the response of the HXC contribution of the PAW effective potential, $V^\sigma_\tr{PAW}$ are added to $V^\sigma_\tr{Hxc}$ in the minimum-tracking definitions of U and J, shown in Eqs. (18a) and (22) of Ref.~\citen{Linscott2018} (where $\sigma=\{\uparrow,\downarrow\}$ is the spin index). Following Refs. \citen{Lambert2023} and \citen{MacEnulty2024}, the responses are fit with polynomial functions of order three (cubic) or lower, the uncertainty of which corresponds to the unbiased standard deviation on the Hubbard parameter (see Appendix A1 for a definition of this uncertainty). The value of the derivative of these polynomials at the molecule's ground-state occupancy (magnetization) is taken as the Hubbard U (Hund's J) parameter. An example of this procedure for [Fe(CO)$_6$]$^{2+}$ can be found in Figure \ref{LR_CO} of Appendix A1.

Global (PBE0\cite{pbe0}) and range separated (CAMB3LYP\cite{camb3lyp}, $\omega$B97X-D\cite{B810189B}) hybrid calculations are performed using ORCA 2.0.3\cite{ORCA}. For all functionals, the relativistic Douglas--Kroll--Hess\cite{Kroll,Hess} Hamiltonian is used, and the calculations are conducted in the unrestricted Kohn-Sham framework (UKS). The basis set aug-cc-pwCVTZ-DK is adopted for iron, while cc-pVTZ is set as default for lighter atoms. The single point calculations are performed under an energy convergence criterium of $10^{-9}$ Ha, and the $\Delta E$ agree with the values from ONETEP (MAE=0.08). Charge density differences are computed using Multiwfn.

\section{Results and discussion}

\subsection{\label{Results_HPs}Linear Response Hubbard Parameters}

The ONETEP minimum-tracking U and J parameters are tabulated in Table \ref{ONETEP_HPs}. All response was well-behaved, reflecting the excellent runtime convergence behavior observed when using the NGWF set provided, resulting in low regression errors across the board, particularly for LS states. 
Noting also that minimum-tracking linear response avoids the need for response inversion, the estimated errors are substantially lower than those sometimes observed in the more commonplace self-consistent field formulation of linear response.

\setlength\arrayrulewidth{1pt}
\setlength{\extrarowheight}{1pt}
\begin{table}[t!]
\footnotesize
\arrayrulecolor{black}
\def\arraystretch{1.4}
\begin{tabular}{ccc|cc|cc}
\multicolumn{1}{l}{}  & \multicolumn{1}{l}{} & \multicolumn{1}{l}{} & \multicolumn{2}{c}{U {\tiny $\pm$ error} [eV]} & \multicolumn{2}{c}{J {\tiny $\pm$ error} [eV]}  \\ \cmidrule(lr){4-5}\cmidrule(lr){6-7}

\multicolumn{1}{c}{} & \multicolumn{1}{c}{Molecule} & \multicolumn{1}{c|}{sub.} & {\color{Sea} HS} & \multicolumn{1}{c|}{\color{Salmon} LS} & {\color{Sea} HS} & \multicolumn{1}{c}{\color{Salmon} LS}  \\ 
\cline{2-7} \multicolumn{7}{c}{}  \\ [-10.7pt]

  & \multicolumn{1}{c}{ \tcw\cellcolor{Barbie} [Fe(CNH)$_6$]$^{2+}$} & Fe $3d$ & \celco{FB9E7B}6.241 {\tiny $\pm\ 0.010$} & \celco{F98458}7.760 {\tiny $\pm\ 0.002$} & \celco{FFFEFE}0.52 {\tiny $\pm\ 0.04$} & \celco{FFFEFD}0.555 {\tiny $\pm\ 4\times 10^{-6}$}  \\
  
  & \multicolumn{1}{c}{ \tcw\cellcolor{OldRose} [Fe(CO)$_6$]$^{2+}$} & Fe $3d$ & \celco{FCB094}5.16 {\tiny $\pm\ 0.14$}  & \celco{FA875C}7.609 {\tiny $\pm\ 2\times 10^{-4}$}  & \celco{FFFEFD}0.556 {\tiny $\pm\ 4\times 10^{-6}$} & \celco{FFFEFD}0.553 {\tiny $\pm\ 2\times 10^{-8}$}  \\
  
  & \multicolumn{1}{c}{ \tcw\cellcolor{Sea} [Fe(NCH)$_6$]$^{2+}$}  & Fe $3d$ & \celco{FBAD90}5.34 {\tiny $\pm\ 0.02$} & \celco{FB9E7B}6.271 {\tiny $\pm\ 0.013$} & \celco{FFFFFE}0.513 {\tiny $\pm\ 9.6\times 10^{-4}$} & \celco{FFFFFE}0.510 {\tiny $\pm\ 5\times 10^{-10}$}  \\
  
\multirow{-4}{*}{\rotatebox{90}{Iron center}} & \multicolumn{1}{c}{ \tcw\cellcolor{Prussian} [Fe(NH$_3$)$_6$]$^{2+}$} & Fe $3d$ & \celco{FCB398}4.999 {\tiny $\pm\ 0.007$} & \celco{FBA787}5.721 {\tiny $\pm\ 0.005$} & \celco{FFFFFE}0.501 {\tiny $\pm\ 4\times 10^{-4}$} & \celco{FFFFFF}0.455 {\tiny $\pm\ 1.3\times 10^{-6}$}  \\ 

\cline{2-7} \multicolumn{7}{c}{} \\ [-10.7pt]

 & \multicolumn{1}{c}{ \tcw\cellcolor{Barbie} [Fe(CNH)$_6$]$^{2+}$}& C $2p$ & \celco{FEE9E1}1.79 {\tiny $\pm\ 0.06$} & \celco{FEE0D4}2.351 {\tiny $\pm\ 1.0\times 10^{-4}$} & \celco{FFFEFD}0.561 {\tiny $\pm\ 3\times 10^{-5}$} & \celco{FFFDFC}0.620 {\tiny $\pm\ 2\times 10^{-6}$}  \\
                            
 & \multicolumn{1}{c}{ \tcw\cellcolor{OldRose} [Fe(CO)$_6$]$^{2+}$} & C $2p$ & \celco{FFEDE6}1.55 {\tiny $\pm\ 0.04$} & \celco{FEE6DD}1.969 {\tiny $\pm\ 3\times 10^{-5}$} & \celco{FFFEFD}0.557 {\tiny $\pm\ 3\times 10^{-5}$} & \celco{FFFDFC}0.614 {\tiny $\pm\ 7\times 10^{-7}$}  \\
                            
& \multicolumn{1}{c}{ \tcw\cellcolor{Sea} [Fe(NCH)$_6$]$^{2+}$} & N $2p$ & \celco{FDC4AF}3.984 {\tiny $\pm\ 0.002$} & \celco{FCBEA7}4.340 {\tiny $\pm\ 7\times 10^{-5}$} & \celco{FFFBF9}0.725 {\tiny $\pm\ 7\times 10^{-5}$} & \celco{FFFAF9}0.755 {\tiny $\pm\ 5\times 10^{-7}$}  \\
                            
\multirow{-4}{*}{\rotatebox{90}{NN}}  & \multicolumn{1}{c}{ \tcw\cellcolor{Prussian} [Fe(NH$_3$)$_6$]$^{2+}$} & N $2p$ & \celco{FCBBA2}4.54 {\tiny $\pm\ 0.02$} & \celco{FCB499}4.918 {\tiny $\pm\ 2\times 10^{-5}$} & \celco{FFF7F4}0.960 {\tiny $\pm\ 7\times 10^{-5}$} & \celco{FFF8F5}0.920 {\tiny $\pm\ 4\times 10^{-6}$}  \\
\cline{2-7} \multicolumn{7}{c}{} \\ [-10.7pt]

 & \multicolumn{1}{c}{ \tcw\cellcolor{Barbie} [Fe(CNH)$_6$]$^{2+}$} & N $2p$ & \celco{FCB69C}4.792 {\tiny $\pm\ 6\times 10^{-4}$} & \celco{FCB195}5.116 {\tiny $\pm\ 3\times 10^{-4}$} & \celco{FFFAF8}0.766 {\tiny $\pm\ 5\times 10^{-6}$} & \celco{FFFAF8}0.766 {\tiny $\pm\ 5\times 10^{-6}$}  \\
 
 & \multicolumn{1}{c}{ \tcw\cellcolor{OldRose} [Fe(CO)$_6$]$^{2+}$} & O $2p$ & \celco{FA9772}6.64 {\tiny $\pm\ 0.02$} & \celco{FA956F}6.782 {\tiny $\pm\ 3\times 10^{-5}$} & \celco{FFF9F6}0.852 {\tiny $\pm\ 2\times 10^{-4}$} & \celco{FFF9F6}0.851 {\tiny $\pm\ 2\times 10^{-7}$}  \\
 
\multirow{-3}{*}{\rotatebox{90}{NNN}} & \multicolumn{1}{c}{ \tcw\cellcolor{Sea} [Fe(NCH)$_6$]$^{2+}$} & C $2p$ & \celco{FEE1D5}2.295 {\tiny $\pm\ 2\times 10^{-4}$} & \celco{FEDBCE}2.610 {\tiny $\pm\ 2\times 10^{-6}$} & \celco{FFFDFC}0.597 {\tiny $\pm\ 5\times 10^{-6}$} & \celco{FFFDFC}0.600 {\tiny $\pm\ 1.1\times 10^{-6 }$}  \\
\cline{2-7} \multicolumn{7}{c}{} \\ [-10.7pt]

 & \multicolumn{1}{c}{ \tcw\cellcolor{Barbie} [Fe(CNH)$_6$]$^{2+}$} & H $1s$ & \celco{FFFAF8}0.778 {\tiny $\pm\ 3\times 10^{-4}$} & \celco{FFF8F5}0.923 {\tiny $\pm\ 0.003$} & \celco{FEE9E1}1.806 {\tiny $\pm\ 3\times 10^{-4}$} & \celco{FEE9E0}1.816 {\tiny $\pm\ 0.002$}  \\
 
 & \multicolumn{1}{c}{ \tcw\cellcolor{Sea} [Fe(NCH)$_6$]$^{2+}$} & H $1s$ & \celco{FFFCFB}0.651 {\tiny $\pm\ 4\times 10^{-4}$} & \celco{FFF9F6}0.855 {\tiny $\pm\ 2\times 10^{-4}$} & \celco{FEE9E1}1.806 {\tiny $\pm\ 8\times 10^{-4}$} & \celco{FEE9E1}1.788 {\tiny $\pm\ 2\times 10^{-4}$}  \\
 
\multirow{-3}{*}{\rotatebox{90}{Hydrogen}}  & \multicolumn{1}{c}{ \tcw\cellcolor{Prussian} [Fe(NH$_3$)$_6$]$^{2+}$} & H $1s$ & \celco{FFF9F7}0.823 {\tiny $\pm\ 3\times 10^{-5}$} & \celco{FFFAF8}0.785 {\tiny $\pm\ 2\times 10^{-5}$} & \celco{FFECE5}1.632 {\tiny $\pm\ 1.4\times 10^{-4}$} & \celco{FFEBE4}1.646 {\tiny $\pm\ 6\times 10^{-5}$}  \\

\end{tabular}
\caption{Site-dependent ONETEP minimum-tracking 
linear response Hubbard Parameters \textrm{U} and \textrm{J} and their regression errors for all spin states ({\color{Sea} HS} or {\color{Salmon} LS}) and all molecular systems, ordered by ligand field strength and grouped in terms of the atomic position in the molecule (\textrm{NN} refers to nearest neighbor; \textrm{NNN} is next-nearest neighbor; sub. refers to treated subspace). Cell color is a function of parameter magnitude relative to all other parameters (i.e., the lighter the orange, the smaller the parameter).}
\label{ONETEP_HPs}
\end{table}

Across all molecules, straightforwardly for the LS and imperfectly for the HS, the Hubbard parameters on the iron center tend to increase with strengthened ligand field. This phenomenon is possibly linked to trends in the Fe magnetic moment, shown in Fig. \ref{hsFeMoment}; the weaker ligand field complexes tend to have larger magnetic moments. It is worth noting that the opposite correlation was observed in Ref.~\citen{MacEnulty2023} for the NiO HPs, where FM NiO had the largest HPs despite harboring the largest magnetic moments. Across all subspaces, it remains probable that trends in the magnitude of the HPs are related to trends in valence occupancy metrics, although the literature on the topic has yet to ascertain the nature of this generally complex and screening-dependent relation.\cite{MacEnultyThesis} The rigidity of the subspace response to a potential perturbation—in other words, the willingness of a subspace to transact with the surrounding electron bath—is a property that's been said to correlate with common chemical properties such as electronegativity\cite{dong2022} and more recently and relatedly in the context of linear response HPs, chemical hardness.\cite{Moore2024} 

In all cases except for that of H on [Fe(NH$_3$)$_6$]$^{2+}$, the U is larger in the LS state as opposed to the HS, in agreement with previous studies.\cite{zhao2016,Mariano2020,Mariano2021,cytter2022} This could derive from the slightly different geometries used for each spin state; Ref.~\citen{Kulik2011} showed that increasing the Fe-O interatomic distance in the FeO$^+$ molecule decreased the value of U. The same trend is present for this series of Fe(II) molecules, for which the HS state has 10\% - 20\% larger metal-ligand bond lengths than in the LS state. Exchange mechanics could also factor into this observation; electrons of like-spin, in experiencing less Coulomb repulsion due to exchange, are more likely to find themselves further from each other. Thus, electrons of like-spin are more delocalized by nature, an effect that is already well replicated by approximate (semi-)local exchange-correlation functionals and therefore demanding of less correction. This lemma is not necessarily reflected in our findings, however; most subspace occupancies (especially the $d$-orbitals of the iron centers, for which the differences between U$_\tr{hs}$ and U$_\tr{ls}$ are greatest) are more integer-like when the molecules are in the LS state. J exhibits very subtle, if any, dependence upon the molecule's spin state. The HPs on the nearest neighbor (NN) carbon follow suit, but not so those for the NN nitrogen, which decrease with respect to increasing ligand field strength.

The magnitude of the Hund's J for the $p$-block elements seems correlated with its uncorrected DFT total occupancy, hovering consistently at around 0.14 - 0.18 eV per electron. Such correlations have not been studied intensively; however, a high-throughput study of Hubbard parameters on transition metal oxides found no exclusive relation between the location of the $d$-block element in the periodic table and its corresponding Hund's J value (see Table 1(b) of Ref.~\citen{Moore2022}). In a separate trend, with the exception of [Fe(NH$_3$)$_6$]$^{2+}$, the further the atom is from the molecular center, the larger the Hund's J.

It is fairly standard across the literature to find a large U value coupled with a small J, which renders the hydrogen U to J ratio of these molecules surprising; the hydrogen J is consistently around twice its U value, i.e., the H-localized static-correlation error
in the approximate functional is around twice the strength
of the H-localized delocalization error. 
Within the standard Dudarev DFT+U$_\tr{eff}$ functional, then, the H $1s$ subspace will receive a negatively valued correction. While the minimum-tracking method does not rely on the unscreened and screened responses to calculate the U (J), we can calculate $\chi$ ($\chi_M$), and with it, reverse engineer $\chi_0$ ($\chi_{0_M}$), using the screened ONETEP occupancy response. With ONETEP, we note that $\chi_0$ and $\chi_{0_M}$ are noticeably dissimilar to each other. This reiterates that the minimum-tracking and self-consistent field linear response methodologies are not equivalent for these molecular systems, by definition, with regards to the unscreened (or internal Kohn-Sham) response.

Another interesting analysis involves focusing on the total of corrective constants for [Fe(NCH)$_6$]$^{2+}$ and [Fe(CNH)$_6$]$^{2+}$, molecules with the same atomic constituents configured slightly differently. The latter seems to demand as much as 2 eV more corrective power. Most valence subspaces see an increase in their parameter values when carbon is closest to the iron center (stronger ligand field). The largest contributor to the discrepancy in corrective power between these two similar molecules comes from the Fe $3d$ subspace, which is accompanied by a small, but noticeable, increase in occupancy. Overall, the reasonability of the ONETEP parameters provides a suitable basis on which to build our energetics and electronic structure investigations.

\subsection{\label{Results_ElectronicStructure}Electronic Structure}

\begin{figure}[t!]
    \centering
    \includegraphics[trim={0cm 0 0.25cm 0},clip,width=1.0\textwidth]{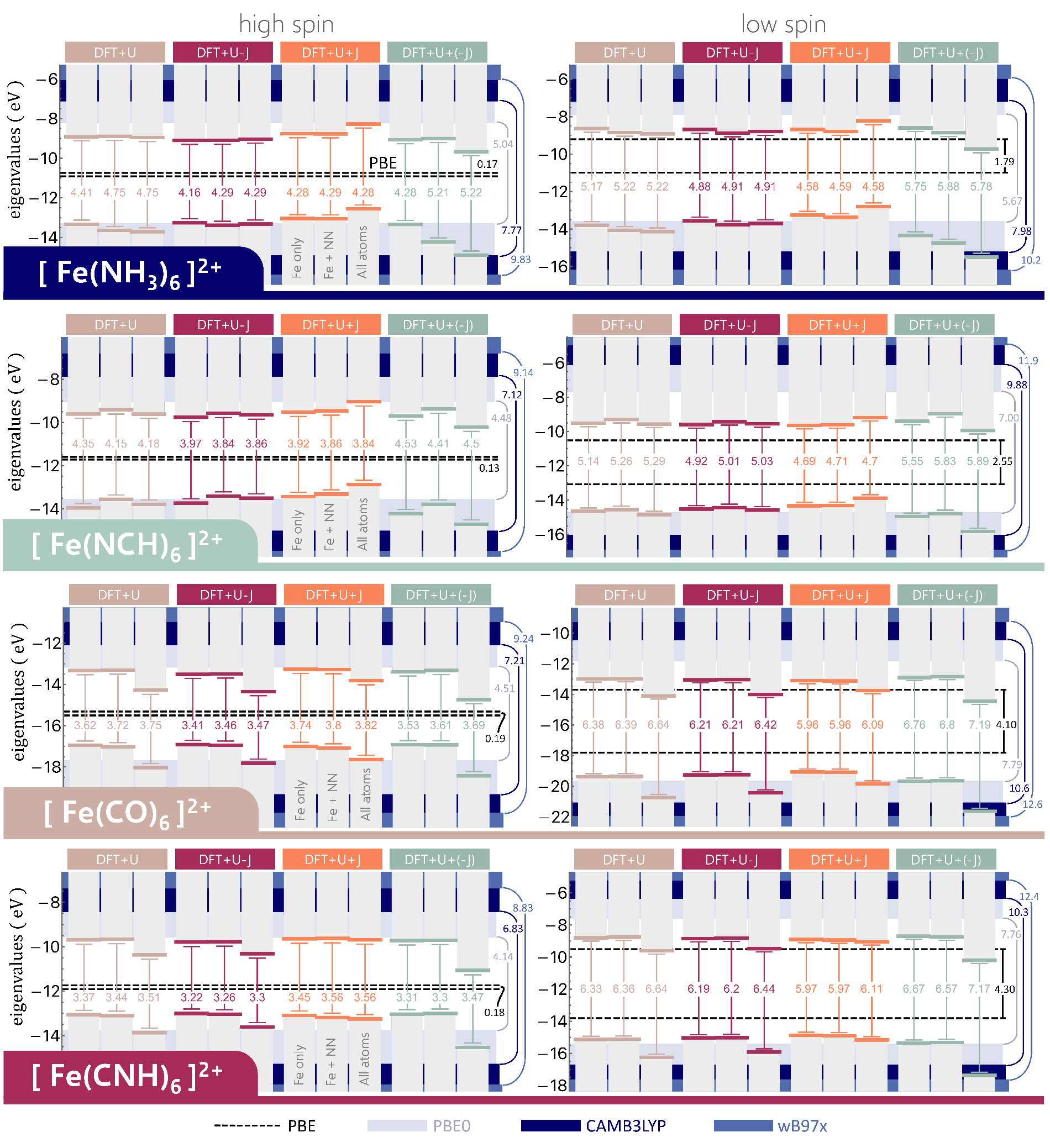} 
    \caption{HOMO and LUMO eigenvalues and gaps (numbers in eV) of all molecules, in HS (left column) or LS (right column) state, as determined by \textrm{PBE} (black dashed lines) and all tested Hubbard functionals (gray, color-lined columns).  PBE0 (light blue platforms), CAMB3LYP (dark blue platforms), and $\omega$B97x (medium blue platforms) hybrid functionals are also shown for comparison. The Hubbard correction is applied, using the \insitu\ Hubbard parameters, to \textrm{Fe} $3d$ alone, to \textrm{Fe} and its immediate neighboring atom $2p$ \textrm{(Fe+NN)}, or on all valence subspaces \textrm{(All atoms)}.}
    \label{BandGaps}
\end{figure}

Any Hubbard functional with \insitu\ correction applied to, at minimum, the Fe $3d$ orbital widens the PBE band gap, according to Fig.~\ref{BandGaps}.

In the LS case, the PBE band gap sensitively increases with increasing ligand field strength, as expected, and the Hubbard U (DFT+U) potential further increases this gap but to an extent that decreases with the ligand's strength. For example, the PBE gap increases from 1.79 eV to 4.30 eV (from NH$_3$ to CNH ligands) and the DFT+U (with U applied to Fe only) increases from 5.17 eV to 6.33 eV, resulting in a gap widened by approximately 2 to 3 eV.
The change in band gap upon Hubbard U correction is mainly attributable to the lowering of the highest occupied molecular orbitals (HOMO), although some changes in the lowest unoccupied molecular orbitals (LUMO) are also visible. This is attributed to the occupation numbers being close to unity for the occupied t$_{2g}$ (HOMO) and more fractional for the e$_g$ (LUMO), resulting in a {\sl strong} attractive Hubbard potential for the former and a {\sl weak} repulsive Hubbard potential for the latter. As the ligand field increases, the occupation numbers of the t$_{2g}$ approach unity, and the Hubbard potential becomes more negative (which can be visualized in Fig. \ref{BandGaps}), thus pushing the HOMO further down.

\begin{figure}[t!]
    \centering
    \includegraphics[trim={0cm 0 0.5cm 0},clip,width=1.0\textwidth]{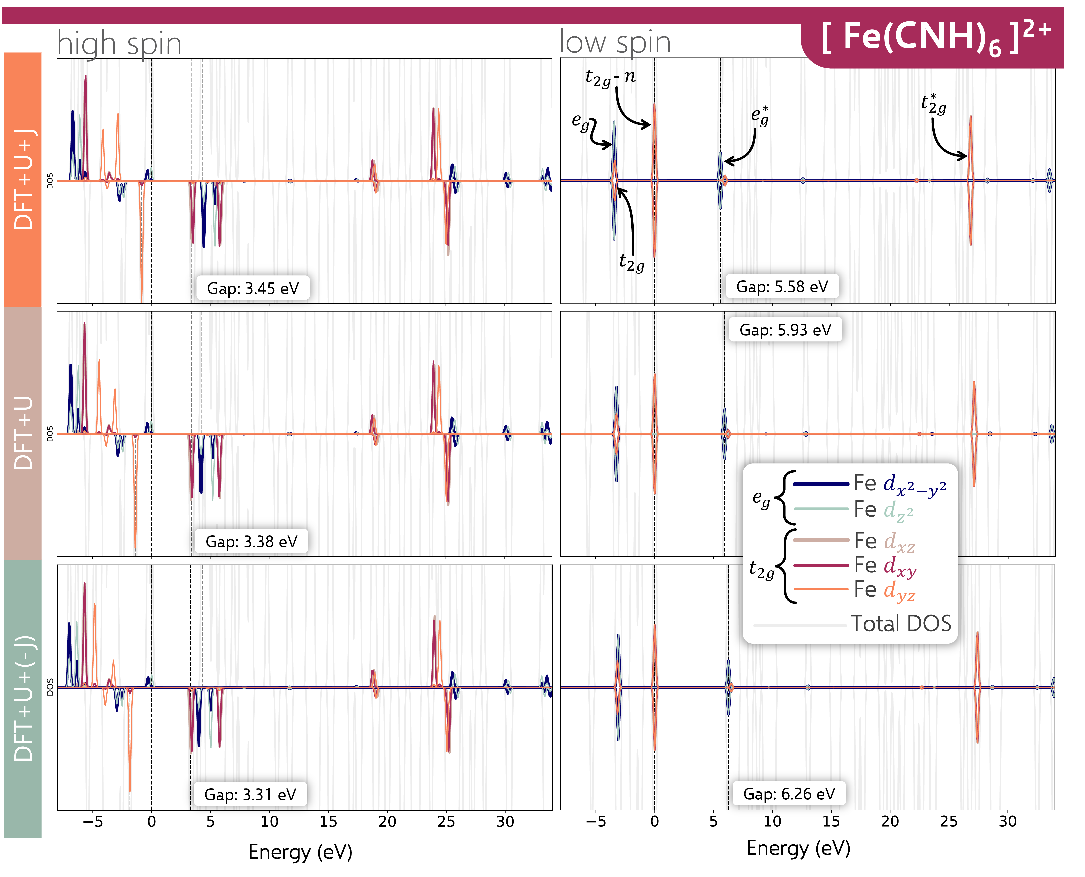} 
    \caption{Projected density of states and HOMO-LUMO gaps for both HS (left column) and LS (right column) for the  [Fe(CNH)$_6$]$^{2+}$ complex using various Hubbard functionals and HS Hubbard parameters (see text). (DFT+U middle row, tan; DFT+U+J top row, orange; DFT+U+(-J) bottom row, green) equipped with HS Hubbard parameter values. The $e_g$, $t_{2g}$, $e^*_g$, $t^*_{2g}$, and $t_{2g}$-n molecular orbitals are labeled in the LS case, where the $e_g$ orbitals comprise the Fe $d_{x^2-y^2}$ (blue) and $d_{z^2}$ (green) orbitals, and $t_{2g}$ comprise the Fe $d_{xy}$ (pink), $d_{xz}$ (tan), and $d_{yz}$ (orange) orbitals. Total DOS is shown in light gray. Dashed black lines indicate frontier (HOMO or LUMO) orbital energies, and for HS the spin-up and spin-down frontier orbital energies are also indicated by light gray dashed lines.}
    \label{PDOS_CNH}
\end{figure}

The source of this behavior is related to how the Hubbard U functional deals with covalency in these molecules. The evolution of the occupancies as a function of ligand field strength is discussed in detail in Ref.~\citen{Mariano2020}, for which we summarize the main points. We recall that, importantly, for the e$_g^*$ orbitals (LUMO), the occupancies are non-zero owing to the occupied ligand-like e$_g$ molecular orbitals, which, at lower energy, exhibit $d$-like character (the projection of the occupied Kohn-Sham states onto a $d$-like atomic basis yields occupancies between 0 and $\frac{1}{2}$) as illustrated in Fig.~\ref{PDOS_CNH}. The weaker the ligand field, the smaller and less fractional these occupancies, since the contribution from the ligand states is lower. Similar arguments hold for the t$_{2g}$; because the t$_{2g}$-like states are mostly occupied, a lower covalency (weaker ligands) results in occupancy values closer to unity, owing to the fact that the unoccupied t$_{2g}^*$ will yield less metallic character (see Fig. 3 of Ref.~\citen{Mariano2020} and related discussion therein).

\begin{figure}[t!]
    \centering
    \includegraphics[trim={0cm 0 0cm 0},clip,width=1.0\textwidth]{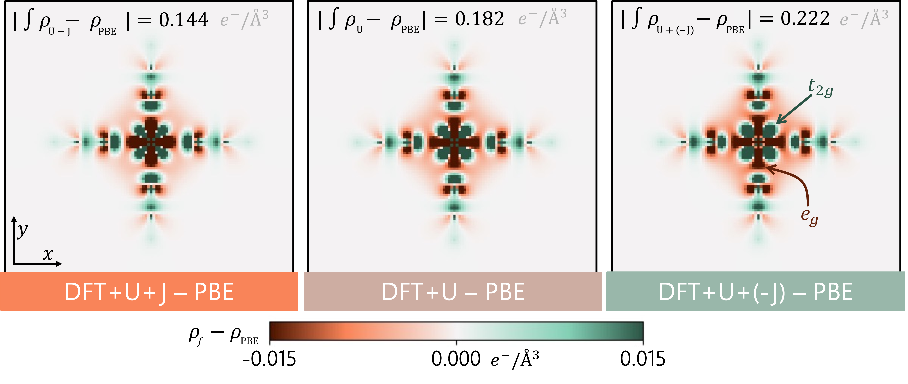} 
    \caption{Charge density difference (in $e^-/\textrm{\AA}^3$) between Hubbard functionals (corrections applied to Fe only) and PBE for [Fe(CNH)$_6$]$^{2+}$ in the LS state (cross-section at $z=10$ \AA, bisecting the iron atom). The densities analyzed are the valence-electron PAW pseudo-densities and do not reflect the all-electron density within the PAW core sphere radii. The absolute value of the integral of that density difference across the entire $20\textrm{\AA}\times20\textrm{\AA}\times20\textrm{\AA}$ cell is written explicitly.}
    \label{Densities_CNH_LS}
\end{figure}

This analysis also illustrates how the Hubbard U potential acts on the Kohn-Sham states and actually changes the metal-ligand covalency. That is, the LUMO is pushed up and the HOMO is pushed down, resulting in a larger (lower) $d$-character for the (latter) former. This can be seen in Fig \ref{Densities_CNH_LS}, where the electronic density difference between DFT+U (U applied to Fe $3d$ only) and PBE is shown for [Fe(CNH)$_6$]$^{2+}$. Negative and positive density differences are found, respectively, for the e$_g$ and t$_{2g}$ orbitals.

In analyzing the case of DFT+U+J with corrections applied to Fe only, the situation is reversed with respect to DFT+U; the DFT+U+J gap increases with respect to PBE (i.e., by 2.79 eV for NH$_3$ and 1.67 eV for CNH), but not as drastically as does DFT+U. Unlike the Hubbard U case, the Hund's J potential is always positive, so the Kohn-Sham states that have a non-zero projection onto the atomic basis are always destabilized, as shown in Fig. \ref{BandGaps}. Thus, both the HOMO and LUMO are upshifted in energy by the Hund's J potential with respect to the U potential, and by a larger extent for the HOMO since the occupancies for the t$_{2g}$ are larger. This results in a sensitive metal-to-ligand charge transfer of t$_{2g}$ symmetry, as illustrated in Fig. \ref{Densities_CNH_LS}.

To summarize the LS case, then, while DFT+U+J inverts the DFT+U correction, DFT+U+(-J) further enhances it because the negative Hund's J potential pushes the HOMO further down in energy with respect to DFT+U, and we thus observe in Fig.~\ref{BandGaps} a gap opening with respect to PBE that is the largest for DFT+U+-J and the smallest for DFT+U+J. This is also noticeable from the charge density plots in Fig. \ref{Densities_CNH_LS}, which show that the integrated charge difference with respect to PBE decreases from DFT+U+(-J) to DFT+U+J.

For HS, the situation is different. First, we notice a small HOMO-LUMO gap in PBE that is very similar for all molecules (around 0.2 eV). The DFT+U functional (with corrections applied to Fe $3d$ orbitals) widens the HS gap more than it does the LS. For example, for NH$_3$, the gap opens by 4.24 eV (versus 3.38 eV for LS), and for CNH it opens by 3.19 eV (versus 2.03 eV for LS). However, because the HS gaps are fairly similar and small for all molecules, and because the width decreases with ligand field strength overall, unlike the LS case, the gap decreases upon application of a U correction. Mutually unlike LS, the unoccupied states are pushed up in energy noticeably. This is because the occupation numbers for the unoccupied states are systematically smaller than for the LS \cite{Mariano2020}, and thus the influence of the Hubbard U potential is larger.

Concerning the role of the Hund's J potential in this case, we see a decrease in the HOMO-LUMO gap from DFT+U+J to DFT+U+(-J) for strong ligand field molecules, as shown in the left-hand panels of Fig.~\ref{BandGaps}. It is particularly instructive to discuss the case of CNH with the support of the PDOS plots in Fig.~\ref{PDOS_CNH}. In this case, the frontier states are of e${_g}$ character for the HOMO and t$_{2g}$ for the LUMO. Because the repulsive Hund's J potential acts on spin-orbitals to a degree that depends on the opposing spin channel's orbital occupancy, the spin-down LUMO is shifted up in energy more than the spin-up e$_g$ LUMO, since the spin-down e$_g$ are mostly unoccupied (see left panels of Fig.~\ref{PDOS_CNH}), thus resulting in a gap opening when the Hund's J is applied (gap opens from 3.37 eV for DFT+U to 3.45 eV for DFT+U+J). Similar arguments can be used to explain the reason for the noticeable decrease in energy difference between the HOMO and the HOMO$-1$ from DFT+U+(-J) to DFT+U+J. Unlike the LS case, the charge density difference with respect to PBE increases from DFT+U+(-J) to DFT+U+J, as illustrated in the Supporting Information (see Figs. S3 and S4).

It is worth noting that the charge density difference between the Hubbard functionals and PBE changes substantially more when the Hubbard corrections are applied to all atoms rather than to Fe only. Despite this, the PDOS seem minimally affected, as shown in Figs. S2 through S4.

Overall, regardless of the Hubbard functional choice, the HOMO-LUMO gap is found to be most comparable to that of PBE0 and smaller than CAMB3LYP and $\omega$B97x, as shown in Fig.~\ref{BandGaps}. Interestingly, we notice in the same figure that Hubbard functionals with corrections applied to all atoms may yield LUMO values lower than PBE, thus bringing into question the utility of applying correction to any subspace beyond the central iron atom.

\begin{figure}[t!]
    \centering
    \includegraphics[width=0.9\linewidth,trim={0cm 0 9cm 0},clip]{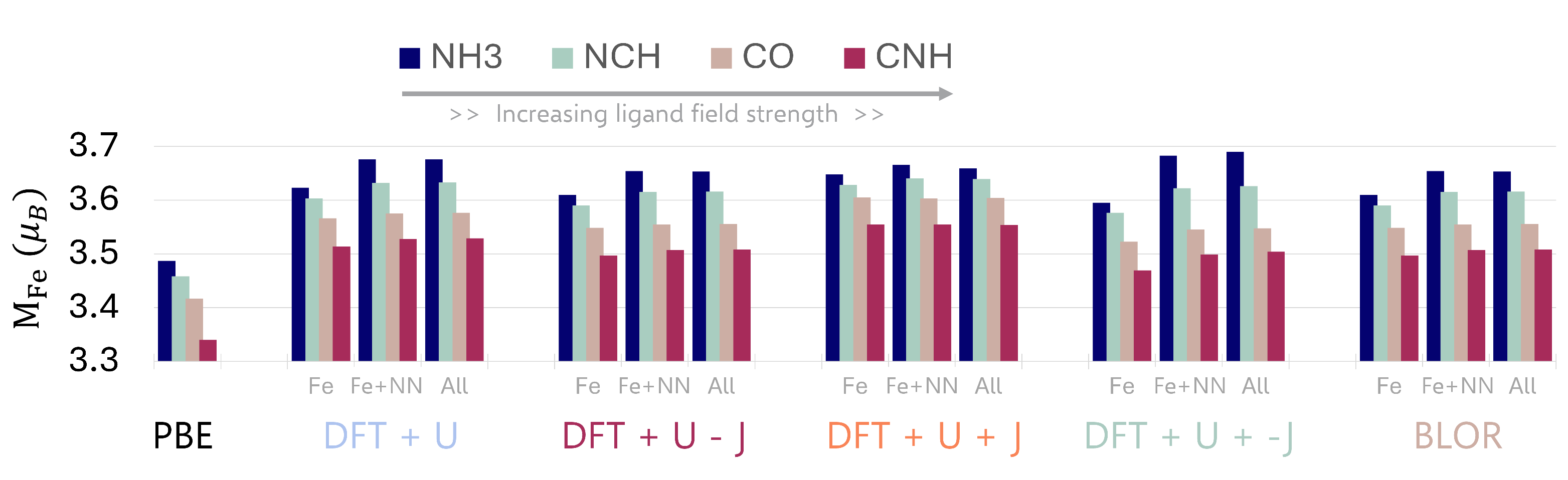} 
    \caption{Magnetic moment for each molecule in its HS configuration as it changes with the Hubbard corrective functional. \textit{In situ} Hubbard corrections are applied to the valence states of the metal only (label: \textrm{Fe}), the metal and the NN (label: \textrm{Fe+NN}), or all atoms in the molecule (label: \textrm{All}).}
    \label{hsFeMoment}
\end{figure}

The magnetic moments predicted in the HS-state coordination of the molecules are shown in Fig. \ref{hsFeMoment}. PBE largely underestimates the magnetic moment with respect to the Hubbard functionals. The moments decrease consistently with increasing ligand field strength, as expected, somewhat linearly and at a rate that depends on the corrective functional. This rate, for example, does not change with a mitigating J term in the Dudarev functional. The rate noticeably decreases when a J correction is added via the Himmetoglu DFT+U+J functional, but increases when the J parameter is made negative. Adding corrections to subspaces beyond the Fe $3d$ is also found to increase the moment.
 
\subsection{\label{Results_Energetics}Energetics}

We compare the energetics deriving from a variety of Hubbard functionals. We present in Table \ref{EnergyDiffsCommon} the adiabatic energy differences \DeltaE\ pertaining to the commonly implemented Dudarev DFT+U$_\textrm{eff}$ = DFT+U-J functional and the Himmetoglu DFT+U+J, as well as the
aforementioned variant of the latter with negatively valued Hund’s J.

\setlength{\extrarowheight}{0pt}
\begin{table}[t!]
\footnotesize
\begin{tabular}{r|cc|cccc|c|c}
\multicolumn{1}{l}{} & \multicolumn{2}{c}{HP specs} & \multicolumn{4}{c}{$\Delta E_\tr{HL}$ (eV)} & \multicolumn{1}{c}{Error (eV)} & \multicolumn{1}{l}{} \\ \cmidrule(lr){2-3}\cmidrule(lr){4-7}\cmidrule(lr){8-8}
Functional & Atoms & \{U, J\} & \cellcolor[HTML]{04017A}{\color[HTML]{FFFFFF} NH3} & \cellcolor[HTML]{A9CDC0}{\color[HTML]{FFFFFF} NCH} & \cellcolor[HTML]{CCADA4}{\color[HTML]{FFFFFF} CO} & \cellcolor[HTML]{A82B5A}{\color[HTML]{FFFFFF} CNH} & & $\overline{\textrm{MAE}}$ \\ \hline
\rowcolor[HTML]{EEF2F4}[\overhang]
CASPT2/CC & -- & -- & -0.64 & -0.16 & 2.02 & 2.87 & \cellcolor{white} & 0.000 \\
\cellcolor{white}{ U   + (-J)} & All & HS & -1.08 & -0.19 & 2.00 & 2.36 &  & 0.25 \\
\cellcolor{white}{ U   + (-J)} & {\color[HTML]{808080} Fe+NN} & HS & -1.26 & -0.20 & 2.34 & 2.45 &  & 0.35 \\
\cellcolor{white}{ U   + (-J)} & {\color[HTML]{BFBFBF} Fe} & HS & -0.77 & -0.40 & 1.90 & 1.96 &  & 0.35 \\
\cellcolor{white}{ U$-$J} & {\color[HTML]{BFBFBF} Fe} & HS & -1.34 & -0.86 & 1.47 & 1.56 &  & 0.82 \\
\cellcolor{white}{ U$-$J} & {\color[HTML]{808080} Fe+NN} & HS & -1.62 & -0.78 & 1.50 & 1.63 &  & 0.84 \\
\cellcolor{white}{ U$-$J} & All & HS & -1.64 & -0.77 & 1.33 & 1.65 &  & 0.88 \\ \hline
\multicolumn{1}{|r|}{\cellcolor{white}{ PBE@U + (-J)}} & {\color[HTML]{808080} Fe+NN} & \cellcolor[HTML]{EDEDEF}\insitu & -0.45 & 0.81 & 3.61 & 4.20 &  & \multicolumn{1}{c|}{1.02} \\
\cellcolor{white}{ U} & {\color[HTML]{BFBFBF} Fe} & HS & -1.49 & -1.06 & 1.20 & 1.30 &  & 1.04 \\
\cellcolor{white}{ U} & {\color[HTML]{808080} Fe+NN} & HS & -1.83 & -0.97 & 1.23 & 1.41 &  & 1.06 \\
\cellcolor{white}{ U} & All & HS & -1.81 & -0.96 & 1.02 & 1.41 &  & 1.11 \\
\rowcolor[HTML]{EEF2F4}[\overhang]
{ PBE } & { --} & { --} & {0.01} & { 1.09} & { 3.72} & {4.21} & \cellcolor{white}{} & { 1.23} \\
\cellcolor{white}{ U   + (-J)} & {\color[HTML]{BFBFBF} Fe} & \cellcolor[HTML]{EDEDEF}\insitu & -1.22 & -0.94 & 0.25 & 1.01 &  & 1.25 \\
\multicolumn{1}{|r|}{\cellcolor{white}{ PBE@U+J}} & {\color[HTML]{BFBFBF} Fe} & HS & 0.05 & 1.18 & 3.89 & 4.48 &  & \multicolumn{1}{c|}{1.38} \\ \hline
\cellcolor{white}{ U+J} & {\color[HTML]{BFBFBF} Fe} & HS & -2.22 & -1.72 & 0.47 & 0.63 &  & 1.73 \\
\cellcolor{white}{ U$-$J} & {\color[HTML]{BFBFBF} Fe} & \cellcolor[HTML]{EDEDEF}\insitu & -1.70 & -1.41 & -0.24 & 0.50 &  & 1.74 \\
\cellcolor{white}{ U+J} & {\color[HTML]{808080} Fe+NN} & HS & -2.35 & -1.74 & 0.11 & 0.36 &  & 1.93 \\
\cellcolor{white}{ U} & {\color[HTML]{BFBFBF} Fe} & \cellcolor[HTML]{EDEDEF}\insitu & -1.83 & -1.61 & -0.50 & 0.26 &  & 1.94 \\
\cellcolor{white}{ U+J} & All & HS & -2.46 & -1.72 & 0.00 & 0.41 &  & 1.96 \\
\cellcolor{white}{ U   + (-J)} & {\color[HTML]{808080} Fe+NN} & \cellcolor[HTML]{EDEDEF}\insitu & -2.70 & -1.23 & -0.72 & -0.15 &  & 2.22 \\
\cellcolor{white}{ U+J} & {\color[HTML]{BFBFBF} Fe} & \cellcolor[HTML]{EDEDEF}\insitu & -2.44 & -2.27 & -1.26 & -0.54 &  & 2.65 \\
\cellcolor{white}{ U   + (-J)} & All & \cellcolor[HTML]{EDEDEF}\insitu & -2.31 & -2.53 & -1.34 & -1.06 &  & 2.83 \\
\cellcolor{white}{ U$-$J} & {\color[HTML]{808080} Fe+NN} & \cellcolor[HTML]{EDEDEF}\insitu & -2.52 & -2.12 & -1.61 & -1.20 &  & 2.88 \\
\cellcolor{white}{ U} & {\color[HTML]{808080} Fe+NN} & \cellcolor[HTML]{EDEDEF}\insitu & -2.63 & -2.36 & -2.08 & -1.64 &  & 3.20 \\
\cellcolor{white}{ U$-$J} & All & \cellcolor[HTML]{EDEDEF}\insitu & -2.34 & -3.39 & -2.04 & -2.00 &  & 3.47 \\
\cellcolor{white}{ U} & All & \cellcolor[HTML]{EDEDEF}\insitu & -2.48 & -3.62 & -2.55 & -2.47 &  & 3.80 \\
\cellcolor{white}{ U+J} & {\color[HTML]{808080} Fe+NN} & \cellcolor[HTML]{EDEDEF}\insitu & -2.54 & -3.46 & -3.42 & -3.11 &  & 4.16 \\
\cellcolor{white}{ U+J} & All & \cellcolor[HTML]{EDEDEF}\insitu & -2.56 & -4.71 & -3.79 & -3.90 & \multirow{-29}{*}{\hspace{-0.0cm}\includegraphics[width=0.199\linewidth,trim={0.6cm 0.1cm 0.36cm 0.145cm},clip]{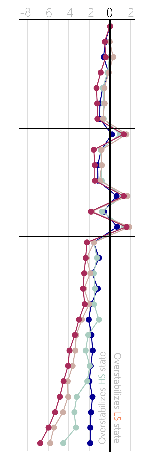}} & 4.76
\end{tabular}
\caption{Ranking of common Hubbard functionals in terms of mean average errors (MAE) for adiabatic energy differences $\Delta E_\tr{HL}$, averaged ($\overline{\textrm{MAE}}$) across all complexes, with respect to \textrm{CASPT2/CC} reference across all molecules (first row).\cite{Mariano2020,Mariano2021} Histogram shows signed error with respect to \textrm{CASPT2/CC} values. \textrm{\{U, J\}} column delineates use of either HS state HP pairs or \insitu\ pairs for each respective spin state. \textrm{“Atoms”} column indicates to which subspaces corrections were applied (\textrm{Fe} = \textrm{Fe} $3d$ only, \textrm{Fe+NN} = \textrm{Fe} $3d$ and \textrm{NN} $2p$, \textrm{All} = all valence subspaces in molecule). Bracketed rows depict the range of average MAE obtained through the density-corrected \textrm{PBE@$f$} functionals (Table S1 of the SI), with the outermost (most and least accurate of the \textrm{PBE@$f$} functionals) given rows themselves.}
\label{EnergyDiffsCommon}
\end{table}

For each corrective functional, we examine the effect of corrective application to select permutations of valence subspaces—to the iron $3d$ alone (Fe), to iron and its neighboring $2p$ (Fe+NN), or to all valence subspaces (All). Then, for each such permutation, we use either the spin-state-specific Hubbard parameters for each spin state (\insitu) or the same Hubbard parameters for both the LS and HS states of the molecules (we opt semi-arbitrarily to use the smaller HS parameters following the success of the DFT+U$-$J functional; \videinfra) the latter method having shown promise for energy differences between magnetic orderings of NiO in Ref.~\citen{MacEnulty2023}. We also include some results from the \emph{density-corrected} functionals, named here PBE@$f$ (labeled PBE[U] in Ref.~\citen{Mariano2021}), with $f$ Hubbard functionals. This approach, as mentioned above, involves removing the Hubbard energy terms after convergence.

\begin{figure}[!ht]
    \centering
    \includegraphics[width=0.76\linewidth,trim={0cm 0cm 0cm 0.2cm},clip]{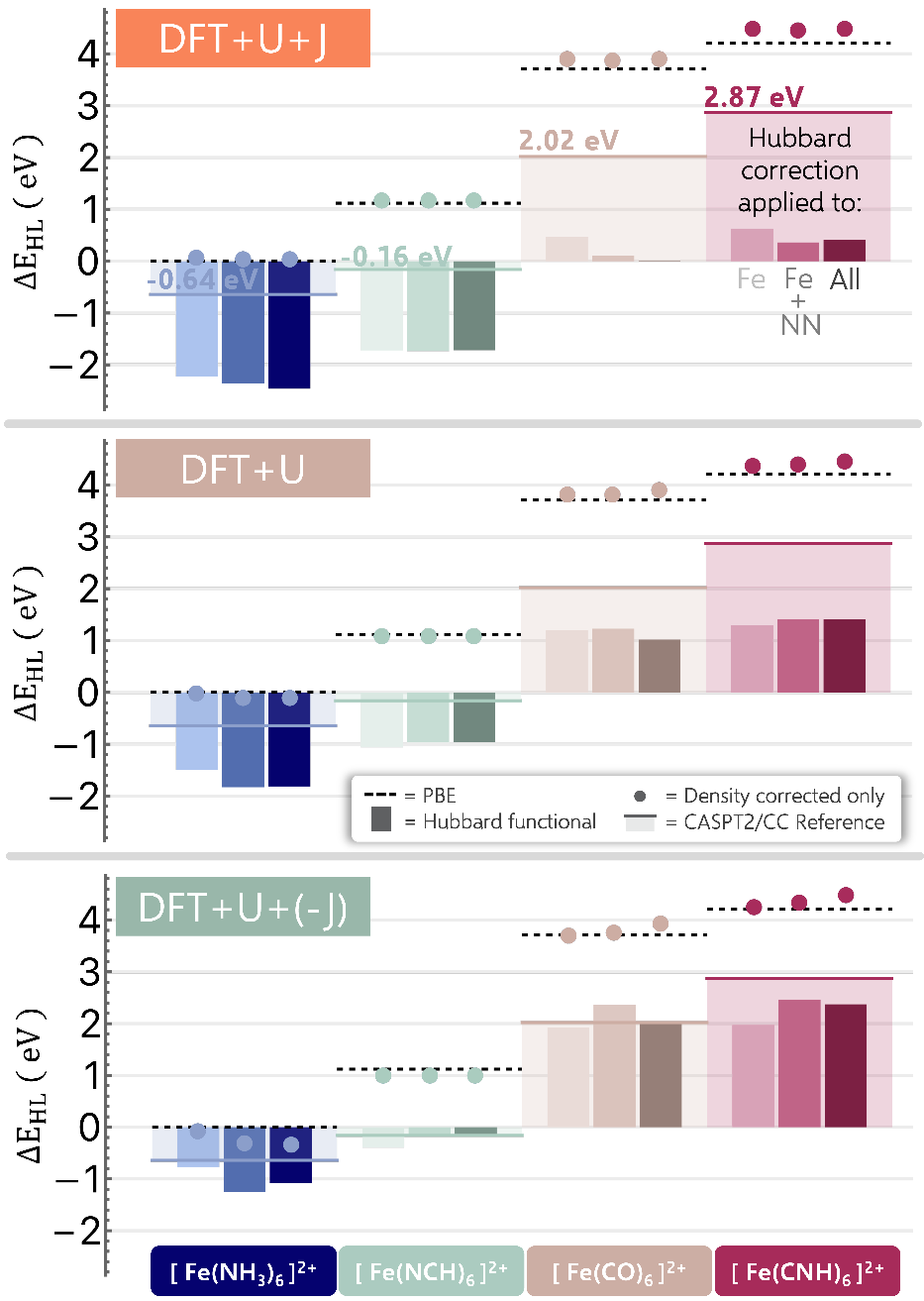}  
    \caption{Adiabatic energy differences \DeltaE\ for all molecules obtained via the denoted Hubbard functional ( Himmetoglu \textrm{DFT+U+J} upper; Dudarev \textrm{DFT+U} middle; Himmetoglu \textrm{DFT+U+(-J)} lower), where correction is applied to some combination of subspaces (\textrm{Fe} = \textrm{Fe} $3d$ only, \textrm{Fe+NN} = \textrm{Fe} $3d$ and \textrm{NN} $2p$, \textrm{All} = all valence subspaces in molecule), calculated with the HS \textrm{\{U, J\}} parameter pairs. Values are compared to PBE (black dashed lines) and \textrm{CASPT2/CC} reference values (solid color line with shading). \textrm{“Density-corrected”} values (data points) are the PBE@$f$ total energy differences converged with the denoted Hubbard functional $f$, but Hubbard energy corrections are removed non-self-consistently.}
    \label{CommonEnergetics}
\end{figure}

Compared to the quantum chemistry reference, uncorrected PBE yields an average MAE magnitude of 1.23 eV across all molecules by overstabilizing the LS state, resulting in positively valued \DeltaE\ across the board. Each functional, including PBE, manages to reconstruct the gradual increase of the adiabatic energy differences with respect to ligand field strength. However, Table \ref{EnergyDiffsCommon} demonstrates that the use of spin-state-specific parameter pairs does not succeed, under any corrective permutation, to outperform PBE; the Hubbard functionals, and in particular the strong field ligand molecules, suffer instead from overstabilization of the HS state. Just as in Ref.~\citen{MacEnulty2023}, this behavior points to a degree of cancellation of errors in those functionals when using the same Hubbard parameter values.

If \insitu\ \{U, J\} pairs are used, however, there is reason to expect an improved density, as it has then been corrected by consistent use of the Hubbard functionals. Using this density correction only, as in PBE@DFT+U (shortened to PBE@U for brevity), yields slightly better \DeltaE\ values than bare PBE if corrections are applied to the iron and the NN only. A comparison of these density-corrected values to the Hubbard functionals themselves can be found in Figs. \ref{CommonEnergetics} and S4 (in the SI), where the PBE@$f$ values for \DeltaE\ for all Hubbard functionals $f$ are similar to those of PBE and thus far from the CASPT2/CC reference values.

We note that this result, particularly concerning those PBE@$f$ values obtained with \insitu\ parameters as displayed in Fig. S4 of the SI, counter those of Ref.~\citen{Mariano2021}, which found that the energetics relative to the same CASPT2/CC reference values are best with PBE[U] and starkly dissimilar to those with PBE. Concerned that this could attest to erroneous execution of our methodology, we performed tests on [Fe(NCH)$_6$]$^{2+}$ to find the source of this discrepancy. We saw no qualitative or otherwise major deviations between reasonably identical PBE and DFT+U (correction on Fe alone) runs between \QE\ (using atomic projectors) and ONETEP. The most notable energetic discrepancy comes from the Hubbard terms ($E_\tr{U,HS}-E_\tr{U,LS}=-2.615$ eV in ONETEP versus -2.384 eV in \QE), which in turn factors dominantly in the discrepancy between the corresponding DFT+U and PBE@DFT+U functionals. Having used the same value of Fe U in our test, this discrepancy of $\sim 0.23$ eV (which increases to $\sim 0.44$ eV when using \emph{ortho}-atomic projectors in \QE, as used in Ref.~\citen{Mariano2021}) attests to the considerable impact of the Hubbard projector function in determining subspace occupancies and derived energies.

We managed, in constructing this experiment, to identify the two most potent differences between our methodology and that of Ref.~\citen{Mariano2021}: (i) the use of atomic-like versus ortho-atomic projectors, and (ii) the use of PAW JTH\cite{Jollet2014} versus GBRV pseudopotentials. Item (ii) is anticipated to account for energy differences on the order of $10^{-2}$ eV (the difference between the PBE \DeltaE\ using JTH versus GBRV pseudopotentials), while energy differences arising from item (i) account for much larger discrepancies, on the order of  $1$ eV. Ortho-atomic or equivalent types of projectors are not implemented in ONETEP at this stage, as the code can handle nonorthogonality easily if needs be. The use of ortho-atomic projectors to find Hubbard subspace occupancies in Ref.~\citen{Mariano2021} is likely the major factor contributing to the fact that their PBE[U] total energies are unlike their raw PBE results. As measures of change in subspace occupancy in and of themselves, the \insitu\ Hubbard parameters are also highly sensitive to the Hubbard projectors.

The best overall option, according to Table \ref{EnergyDiffsCommon}, is to use the same HPs regardless of spin state, at least using 
contemporary conventional Hubbard functionals. In availing of what appears to be a black-box cancellation of errors when the HPs are the same for both HS and LS, in the structure of the Hubbard functional, the overstabilization of the HS state is mitigated and we observe most Hubbard functionals making a decent improvement on PBE. The best-performing physically derived Hubbard functional is the Dudarev DFT+U$_\tr{eff}$, which reduces the average MAE by 34\% with respect to PBE, with correction applied exclusively to Fe $3d$. The use of the J in mitigating the magnitude of U on any subspace is beneficial here, since DFT+U (without J) yields a 16\%-only improvement on PBE. This shows how the large values of U prescribed by linear response are responsible for the overstabilization of the HS states. Incidentally, this also highlights the reason why we selected the HS HP parameters, as they are smaller than both their LS counterparts and hence also than the average of the LS and HS parameters.\footnote[4]{Assuming the density itself is rather well-corrected at its base and thus unperturbed by small changes in the magnitude of the Hubbard parameter, we tested non-self-consistently different \{U, J\} pairs on the adiabatic energy differences. The set of parameters obtained by averaging those of the HS and LS states yielded slightly worse MAEs, as did the LS state set of parameters. To confirm, we applied the HS parameters via the Hubbard functionals self-consistently, however, to obtain the data represented in Table \ref{EnergyDiffsCommon}.} The results are clear in that for this test set, there is no value in applying correction to any subspace beyond the iron valence, although doing so is unlikely to change the adiabatic spin-flip energy differences drastically.

For convenience in the remaining discussion, we rephrase the energy correction on subspace $i$ (dropping the $i$ superscript for brevity) for the Himmetoglu DFT+U+J functional in Eq. \ref{HimmetogluDFT+U+J} as a sum of $m$-and spin-resolved occupancy combinations prefixed by U and J respectively,
\begin{align}\label{HimmetUJTerms}
E_{\U+\J}&=E_\U+E_\J=\frac{\U}{2} \Sigma_\U+\frac{\J}{2} \Sigma_\J,
\end{align}

\noindent where $\Sigma_\U=\sum^{\uparrow,\downarrow}_{\sigma}\sum_{m} n^\sigma_{mm}-(n^{\sigma 2})_{mm}$ and $\Sigma_\J=\sum^{\uparrow,\downarrow}_{\sigma}\sum_{m m'} n^\sigma_{mm'}n^{\bar{\sigma}}_{m'm}-\Sigma_\U$. Reformulating Eq. \ref{HimmetogluDFT+U+J} in this manner permits us to isolate the effect of $E_\J$ in modifying the underlying DFT+U functional.

In Fig.~\ref{CommonEnergetics} and Table \ref{EnergyDiffsCommon}, we see that DFT+U+J egregiously undershoots the target CASPT2/CC reference for each molecule, and the best agreement with the benchmark is obtained when the sign of $E_\J$ is reversed, i.e., in the DFT+U+(-J) functional.
More specifically, overall, while DFT+U+J further destabilizes the HS state with respect to PBE, DFT+U+(-J) does the opposite, as expected by construction since the Hund's J term discourages anti-aligned spins (see Eqs. \ref{HimmetogluDFT+U+J} and \ref{Himmetoglupot}). This is visually clear when looking at the comparison in Fig.~\ref{CommonEnergetics}, where only results obtained using HS Hubbard parameters are shown for clarity. (The same plot for \insitu\ HP energetics can be found in Fig. S4 of the SI).  Under DFT+U+(-J) with HS parameters, the molecules show no particular tendency to overstabilize the low- or high-spin states, and where there is bias, it is not correlated with ligand field strength (see top rows in Table~\ref{CommonEnergetics}). These properties of DFT+U+(-J)—in addition to the fact that the best \DeltaE\ are obtained when correction is applied to all valence subspaces in the molecule, not just the Fe $3d$—are indications that the incorporation of intra-atomic exchange in the Himmetoglu functional may be problematic for isolated 
systems without (in the physical world) spontaneous spin-symmetry breaking, and that the structure of the Hubbard functional itself warrants revision, as already undertaken elsewhere.~\cite{burgess2023,burgess2024}

\begin{figure}[t!]
    \centering
    \includegraphics[width=1.0 \linewidth,trim={0.2cm 0.6cm 0.5cm 0cm},clip]{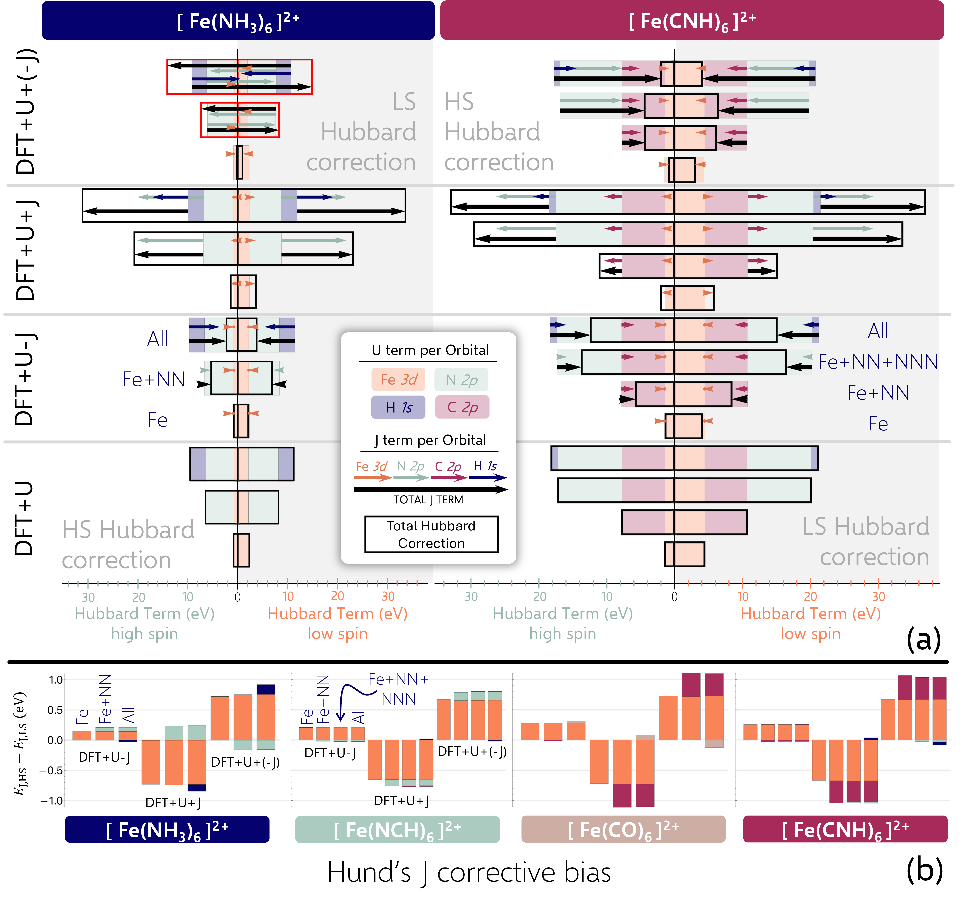} 
    \caption{Corrective energy terms and the Hund's J terms' contribution to the bias in \DeltaE. The plot in (a) splits the Hubbard correction into terms preceded by the Hubbard \textrm{U} (colored, stacked bars), Hund's \textrm{J} (colored arrows; total \textrm{J} correction, which amounts to the sum of the colored arrows, is denoted by black arrow), and then the total of the two (black-rimmed rectangles) for (top left) \textrm{[Fe(NH$_3$)$_6$]$^{2+}$} and (top right) \textrm{[Fe(CNH)$_6$]$^{2+}$} across all Hubbard functionals. Corrections applied to the \textrm{Fe} $3d$ subspace is denoted by orange hues, \textrm{N} $2p$ by green hues, \textrm{C} $2p$ by pink hues, and \textrm{H} $1s$ by blue hues. Red-rimmed rectangles signal that total Hubbard correction applied via that functional is negatively valued. The left side of each paired bar chart is the HS state of the molecule, and the right is the LS. Plot in (b) shows the Hund's \textrm{J} corrective bias.}
    \label{JBias}
\end{figure}

As found with prior investigations, the tendency for $E_\U$ to overstabilize the HS state comes from an inflated penalty applied to the LS state,\cite{Mariano2020} rendering $E_\tr{U,HS}-E_\tr{U,LS}$ negatively valued in a manner increasing in magnitude as one moves to the right of the spectrochemical series. In Fig. \ref{JBias}(a), we see that $E_\U$ for each valence subspace stays approximately the same magnitude regardless of the functional used, where the bias in penalizing the LS state comes primarily from the Fe $3d$ term. It's the J term that changes noticeably with respect to spin state for all subspaces, particularly on the Fe $3d$ and the NN $2p$. These changes accumulate and manage to reduce the penalty bias against the LS state, amplify it, or augment the penalty bias against the HS state.

From Fig. \ref{JBias}(b), the dominant contribution from the Fe $3d$ on the $E_\J$ term becomes clear. Furthermore, the plots illustrate the necessity of a negative $E_\J$ in reducing the bias against the LS state.  The N and C $2p$ corrections, largely canceling themselves out when not immediately neighboring the Fe atom, only contribute anywhere from 25\%-50\% of the Fe $3d$ bias, the direction of which is highly dependent on the functional used. For example, in in [Fe(NH$_3$)$_6$]$^{2+}$, the J correction on N bolsters whatever bias the $3d$ orbital demonstrates in DFT+U$-$J, but mitigates it in the Himmetoglu functional (and its -$E_\J$ variant), only to be canceled, in part or in full, by a H $1s$ J correction almost always demonstrating the opposing bias. The only time the $1s$ and N $2p$ corrections compound their bias is in the DFT+U+(-J) functional on [Fe(CNH)$_6$]$^{2+}$, where they encourage LS bias, rather futilely considering that together they make up less than 10\% of the $3d$ and C $2p$ correction that collectively penalize the HS state more than the LS. Coincidentally, the aforementioned N $2p$ is the only subspace whose correction doesn't flip its bias going from DFT+U+J to DFT+U+(-J); in both functionals, the N contributes to the bias towards the LS state. Coincident to that, the [Fe(CNH)$_6$]$^{2+}$ N $2p$ is the only subspace of all the molecules tested for which the \insitu\ J value is the exact same in both the HS and LS states. The N $2p$ subspace in [Fe(CNH)$_6$]$^{2+}$ is highly spin-polarized and almost fully occupied, featuring the second largest J-scaled energy correction of any subspace (the first being the N $2p$ in [Fe(NH$_3$)$_6$]$^{2+}$); the spin state does not alter this much at all. What's significant is that it is the only subspace for which the $m$- and $\sigma$-resolved occupancy sum $\Sigma_\tr{J,HS}-\Sigma_\tr{J,LS}$—a term defined expressly in Eq. \ref{HimmetUJTerms} to be independent of the sign of the J parameter—switches sign anyway when a negative Hund's J parameter is used as opposed to its standard positive value. That is, in DFT+U+J, the HS $\Sigma_\J$ is smaller than the LS $\Sigma_\J$, whereas in DFT+U+(-J), the opposite is true. We reason through the cause of this behavior in Appendix A2.

What's also interesting, on both [Fe(CNH)$_6$]$^{2+}$ and [Fe(CO)$_6$]$^{2+}$, is that the C $2p$ corrective bias noticeably lessens when corrections are applied to their outer $2p$ neighbor (either N or O). This reflects a larger difference in $\Sigma_\J$'s same-spin penalty $\sum^{\uparrow,\downarrow}_{\sigma}\sum_{m} n^\sigma_{mm}n^{\bar{\sigma}}_{mm}$ between the HS and the LS states; this same-spin penalty reduces for both spin states, but the LS faster than the HS. Because the magnetic moment in the LS state of these molecules is not increasing, this suggests that the J correction on the neighboring $2p$ is causing more of its charge to transfer, in equal parts spin-up and spin-down, to the C $2p$. This is a testament to how much the J correction is affected by the spin degree of freedom, not necessarily the magnitude of the magnetic moment. The strong covalency of the systems perhaps amplifies this.

Across all molecules, the DFT+U$-$J functional lightly counters the larger LS U penalty with a larger J correction on the HS $3d$ orbital, a bias toward the HS state often minimally mitigated by the C $2p$ correction only to be lightly bolstered by other subspace corrections. By contrast, the C $2p$ correction compounds the bias of the Fe $3d$ corrections in the Himmetoglu-type functionals. With DFT+U+J, that bodes poorly for the energetics; the main J correction greatly amplifies the LS state bias, pushing the total energy further away from the CASPT2/CC reference. This is precisely the behavior that is flipped on its head with DFT+U+(-J); the difference in correction largely remains the same magnitude for $3d$ and C $2p$, but it administers the penalty to the HS state instead of the LS, thereby countering the LS bias in DFT+U and resulting in adiabatic energy differences more in line with the CASPT2/CC expectations.

\section{\label{SCOconclusion}Summary and conclusions}

Whether the electronic structure and energetics of SCO complexes is a realm accessible to density functional approximations using semi-local functionals is still an open question. Building on literature in the area, this investigation sought to unearth the fine details of fully first-principles Hubbard-like DFT+U+J methods and their potential to achieve high-precision adiabatic energy differences.

We calculated and analyzed trends of the minimum-tracking linear response-derived Hubbard U and Hund’s J for all valence subspaces in a series of highly covalent, octahedrally-coordinated Fe(II) SCO molecules, adopting either the $^1A_{1g}$ low-spin or $^5T_{2g}$ high-spin state, spanning the ligand field strength spectrum. Having calculated the HPs with ONETEP, we methodically applied them via a select range of common Hubbard functionals in search of the simplest combination to yield reliable spin-state energetic properties with respect to those obtained by our chosen reference: the CASPT2/CC wavefunction method. A brief check on electronic structure properties revealed anticipated corrections to the density, which gave us ground on which to build an energetics analysis. Following this, we found a somewhat counterintuitive failure of the \insitu\ HPs, hinting at a breakdown in fortuitous cancellation of error when the same parameters are used for different spin states. In particular, as motivated from the recent BLOR functional and as verified by experimental DFT+U+(-J) calculations (intended as enabling a proof-of-principle and not proposed here as a functional for any further use), we found that use of the conventional positively valued Hund’s J term in DFT+U+J fails 
in furthering DFT+U’s already robust capacity to obtain reasonable adiabatic energy differences via the Dudarev functional. We explained contradictions in our results with respect to those obtained previously in Refs.~\citen{Mariano2020} and \citen{Mariano2021}, suggesting that the value of PBE@$f$-type density-corrected functionals can be useful depending on the type of projector used. Similar to the conclusions of Ref.~\citen{MacEnulty2023}, it appears best practice (at least when using the currently well-established Hubbard-model rather than flat-plane based functionals) to use the same Hubbard parameter values regardless of the molecule's spin state.

There is a tendency for Hubbard functionals to more strongly penalize the LS state as opposed to the HS state, the opposite trend to that seen when using hybrid functional corrections. The DFT+U+J approach further enhances the trend already observed for DFT+U functionals. Ultimately, however, this investigation supports the case for the construction of more appropriate DFT+U+J-type functionals to account for the  static-correlation phenomena at play in strongly covalent systems. We refer the reader to Ref.~\citen{burgess2023} for a discussion on the BLOR functional, which may provide some insight into why the J term in the Hubbard-model DFT+U functionals to date, needs to be different, and not simply through a change in the sign of the term that it pre-multiplies.

Along the way, we have identified  simple  systems for which first-principles-parameters DFT+U+J breaks down for the energetics, and in doing so, we were able to map previously uncharted limitations of the method and precisely highlight the areas for improvement. As a result, we have reduced the search space across Hubbard-like methodologies, which is a necessary preliminary step to tackling more complicated analogous systems, for example, Prussian Blue and its analogues. Future work and extensions to this project, then, could involve applying these conclusions to ferrous-hexacyanometallate systems for which the discussed molecules are localized and periodically repeating constituents. Getting a DFT+U+J type approach to work on such challenging systems would amount to considerable progress in computationally feasible materials simulation.

\begin{acknowledgement}

The authors thank l'Ambassade de France en Irlande for providing, via the France-Ireland Research Residency scheme of 2023, a basis on which we could launch this collaborative work. LM and DDO acknowledge the Trinity College Dublin Provost PhD Project Awards. The work was further supported by Taighde Éireann — Research Ireland [19/EPSRC/3605], the Engineering and Physical Sciences Research Council [EP/S030263/1]. This publication has emanated from research conducted with the financial support of Taighde Éireann — Research Ireland, Grant Number 12/RC/2278\_2, and is co-funded under the European Regional Development Fund under the AMBER award. Calculations were performed on the Boyle cluster, funded through grants from the European Research Council and Taighde Éireann and maintained by Research IT at Trinity College Dublin. [Insert RP and JPAdM acknowledgements.]

\end{acknowledgement}

\section{\label{Appendix} Appendix}
\appendix

\subsection{\label{SCOlr}\textbf{A1.} Linear response calculations\protect}

To calculate the Hubbard U and the Hund's J, we employ the minimum-tracking linear response definitions by applying a perturbing potential $d\hat{V}_\tr{ext}=dV^i_\tr{ext}\hat{P}^i$ to the Hubbard subspace $i$ and recording the response of the Kohn-Sham potential of the subspace, finding
\begin{equation}\label{UJwspin}
\textrm{U}=\frac{1}{2}\frac{d (V^\uparrow_\textrm{hxc}+V^\downarrow_\textrm{hxc})}{d(n^\uparrow+n^\downarrow)}\ \ \ \ \ \ \ \ \ \ \ \ \ \textrm{J}=-\frac{1}{2}\frac{d (V^\uparrow_\textrm{hxc}-V^\downarrow_\textrm{hxc})}{d(n^\uparrow-n^\downarrow)}\ ,
\end{equation}
where $n^\uparrow \pm n^\downarrow=\textrm{Tr}[ \hat{P}(\hat{\rho}^\uparrow \pm\hat{\rho}^\downarrow)]$ is the total charge occupancy $N$ / magnetization $M$ as a function of the spin density operator $\hat{\rho}^\sigma$, and $V^\sigma_\textrm{hxc}$ is the Hartree + exchange correlation potential for spin $\sigma$. An example input file for performing minimum-tracking linear response in ONETEP is provided in the SI. We refer the reader to Ref.~\citen{Linscott2018} for more information on spin considerations in this methodology. In order to include the response of the HXC contribution of the PAW effective potential, $V^\sigma_\textrm{PAW}$ have been added to $V^\sigma_\textrm{hxc}$ in Eq. \ref{UJwspin}.

\begin{figure}[ht]
    \centering
    \includegraphics[width=0.9\linewidth]{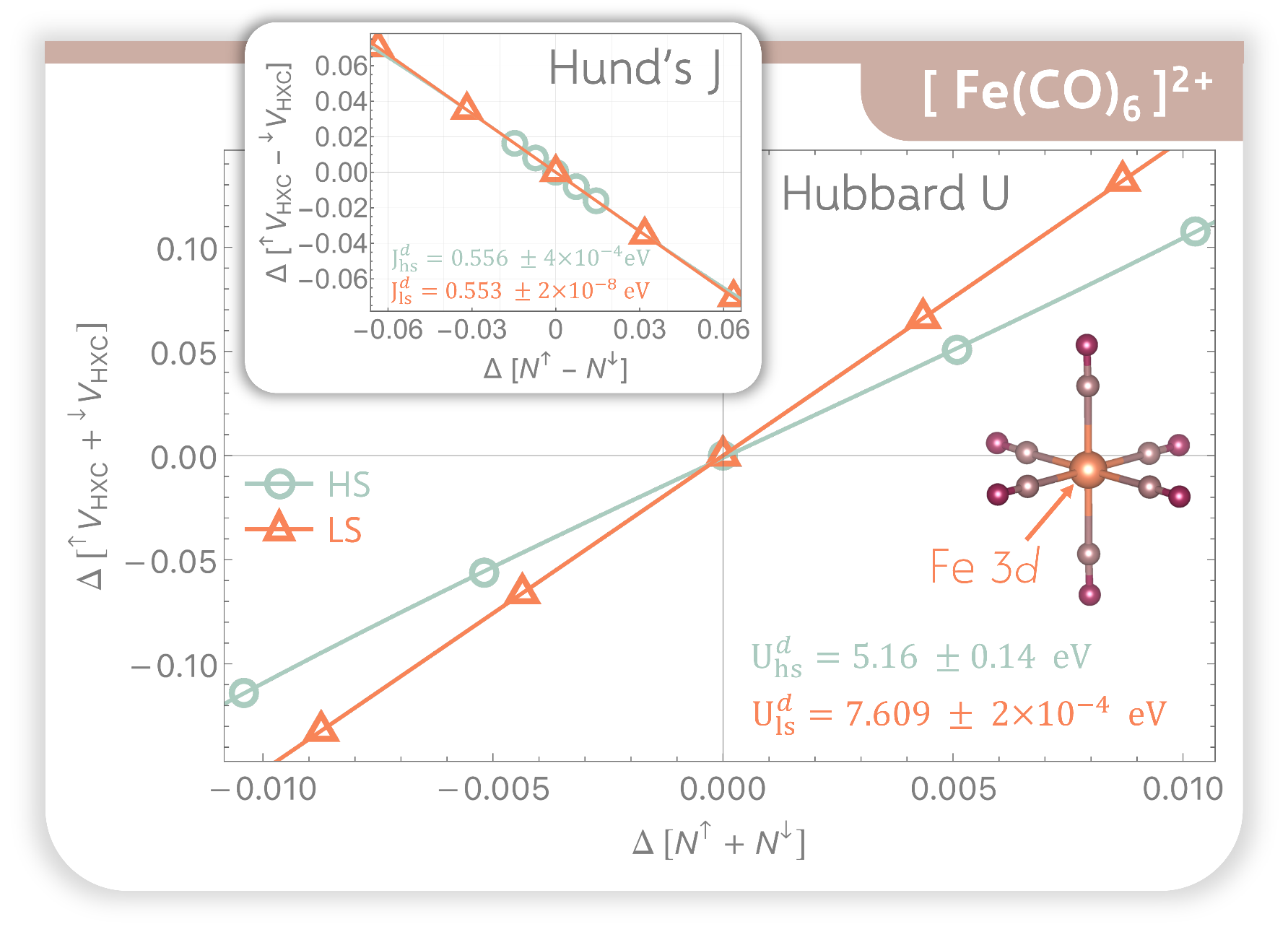} 
    \caption{Example of minimum-tracking linear response conducted on the $3d$ orbitals of the central \textrm{Fe(II)} ion of HS- and LS-coordinated \textrm{[Fe(CO)$_6$]$^{2+}$} to determine the Hubbard \textrm{U} and the Hund's \textrm{J} (inset) parameters.}
    \label{LR_CO}
\end{figure}

We address non-linearity in the response by fitting degree $p$ polynomial regressions $f^{(p)}\left(N(\alpha)\right)=\sum^{p}_{q=0} c_{q} N(\alpha)^{q}$, where $c_{q}$ are the polynomial coefficients determined via a least-squares fitting to the response data set. In the case of the Hubbard U, the response data comprises $m$ occupancy-HXC potential pairs $\left( N(\alpha_i), V^+_{\textrm{hxc}_i} \right)$, where $N(\alpha_i)=\left[ n^\uparrow + n^\downarrow \right] \big\rvert_{\alpha_i}$ is the total occupancy arising from the $\alpha_i$-perturbed subspace, and $V^+_{\textrm{hxc}_i}=V^\uparrow_{\textrm{hxc}_i} + V^\downarrow_{\textrm{hxc}_i}$. The Hubbard U is then calculated through the evaluation of the derivative of the regression at $\alpha=0$,
\begin{equation}\label{HPpolynomial}
\textrm{U}=\frac{1}{2}\sum^{p}_{q=0} q\ c_q\ N(0)^{q-1} \ .
\end{equation}
The Hubbard U is thus a multivariate function with respect to the fitted coefficients. It is important (and, indeed, numerically imperative) to assert that the fitted polynomial coefficients are covariate, meaning their uncertainties do not vary independently. Therefore, the error on the minimum-tracking LR Hubbard U is found to be
\begin{equation}\label{HPerrorsMT}
\sigma^{(p)}_\textrm{U}=\frac{1}{2} \sqrt{ \sum^{p}_{q=0} \sum^{p}_{r=0} q\ r\ C_{q+1,r+1}\ N(0)^{q+r-2}} \ \ \ ,
\end{equation}
where we use the unbiased standard deviation and the $m \times (p+1)$ design matrix $\mathbf{A}$ with elements $A_{i,q+1}=N(\alpha_i)^{q}$ to compute the covariance matrix $\mathbf{C}$, featuring matrix elements
\begin{equation}\label{Covariance}
C_{q+1,r+1} = \frac{ \sum^m_{i=1} \left[ V^+_{\tr{hxc}_i} - f^{(p)}\left(N(\alpha_i)\right) \right]^2 }{m-p-1} \left( \mathbf{A}^\top \mathbf{A} \right)_{q+1,r+1}^{-1} \ .
\end{equation}

Uncertainty on the Hund's J may be ascertained analogously by replacing all instances of $\alpha$ with $\beta$, total occupancy $N(\alpha)$ with subspace magnetization $M(\beta)=\left[ n^\uparrow - n^\downarrow \right] \big\rvert_{\beta}$, and $V^+_{\textrm{hxc}_i}$ with $V^-_{\textrm{hxc}_i}=V^\uparrow_{\textrm{hxc}_i} - V^\downarrow_{\textrm{hxc}_i}$. Furthermore, the $\frac{1}{2}$ prefactor in Eq. (\ref{HPpolynomial}) should be replaced by $-\frac{1}{2}$. Encouragingly, for all subspaces on which we conduct MT linear response in this investigation, we found that the uncertainty incorporating the covariance between polynomial coefficients, described by Eqs. (\ref{HPerrorsMT}) and (\ref{Covariance}) is, in practice, reasonably identical to the regression error obtained when shifting the $N(\alpha_i)$ values by $-N(0)$. In this case, one may evaluate the derivative about a zero-perturbation axis, rendering the HP a singly-variate function of $c_1$. Put more plainly, if one shifts the $N(\alpha_i)$ values by $-N(0)$ before regression is
performed, then one only needs to be concerned with the error in the coefficient $c_1$.

We emphasize here that quantification of the uncertainty on the Hubbard parameters, specifically those arising from response demonstrating non-linear behavior, is a topical research query that warrants more consideration than is given in this article. The definition and appropriateness of the unbiased standard deviation in the response context, for example, is not a universally agreed-upon matter. The application of state-of-the-art statistical techniques to linear response merits its own systematic investigation that lies beyond the scope of this article.

\subsection{\label{Appendix_N2poccs}\textbf{A2.} Nitrogen $2p$ occupancies in [Fe(CNH)$_6$]$^{2+}$\protect}

The N $2p$ subspace in [Fe(CNH)$_6$]$^{2+}$ features the second largest J-scaled energy correction of any subspace (the first being the N $2p$ in [Fe(NH$_3$)$_6$]$^{2+}$); the spin state does not alter this much at all. It is also the only subspace for which the $m$- and $\sigma$-resolved occupancy sum $\Sigma_\tr{J,HS}-\Sigma_\tr{J,LS}$—a term defined expressly in Eq. \ref{HimmetUJTerms} to be independent of the sign of the J parameter—switches sign anyway when a negative Hund's J parameter is used as opposed to its standard positive value. That is, in DFT+U+J, the HS $\Sigma_\J$ is smaller than the LS $\Sigma_\J$, whereas in DFT+U+(-J), the opposite is true.

To understand why this is happening, we look at the six sets of N $2p_x$, $2p_y$, and $2p_z$ occupancies in [Fe(CNH)$_6$]$^{2+}$ (18 orbitals in total) for each functional. For all orbitals, DFT+U+(-J) renders larger spin-up and spin-down occupancies than DFT+U+J, especially for those orbitals that lie off the bond axes. This makes the on-axis contributions to $\Sigma_\J$ 50-60\% larger than those from the off-axis orbitals (for DFT+U+J, for example, the average $\Sigma_\J$ for on-axis orbitals is 1.719 in the HS and 1.717 in the LS, compared to 1.063 and 1.068 respectively for the off-axis orbitals).\footnote{It is possible to identify these contributions by reformulating Eq. \ref{HimmetUJTerms} in terms of operations between on-diagonal occupancy matrix elements (where $m'=m$) plus a second-order correction comprising operations between off-diagonal occupancy matrix elements (this correction is small because the off-diagonal elements of the N $2p$ occupancy matrices are very small).} But the on-axis $\Sigma_\J$ contributions are comparatively resilient to changes in spin state; they typically more heavily penalize the HS state, but that bias is usually 1-2 orders of magnitude weaker than the off-axis contributions. Thus, the $E_\tr{J,HS}-E_\tr{J,LS}$ term we see in Fig. \ref{JBias}(b) mainly comprises contributions from the off-axis orbitals. It is this dynamic that results in the aforementioned phenomenon. In DFT+U+J, most off-axis orbitals not only penalize the LS state but do so strongly, rendering $\Sigma_\tr{J,HS}-\Sigma_\tr{J,LS}<0$. By contrast, in DFT+U+(-J), only half of the off-axis orbitals manage only to weakly penalize the LS state, rendering $\Sigma_\tr{J,HS}-\Sigma_\tr{J,LS}>0$.

\begin{suppinfo}

Example minimum-tracking linear response ONETEP input file; table ranking average MAE across all molecules for PBE@$f$ (density-corrected) Hubbard functionals; charge density differences of all Hubbard functionals (HS Hubbard parameters on iron $3d$ only and then on all valence subspaces) with respect to PBE for HS and LS [Fe(CNH)$_6$]$^{2+}$; figure demonstrating performance of the performance of Hubbard functionals incorporating \insitu\ Hubbard parameters.

\end{suppinfo}

\bibliography{Manuscript}

\end{document}


\subsection{\label{SI_SampleInput}Sample minimum-tracking linear response ONETEP input file}

\noindent Linear response J perturbation of strength $\beta=-0.05$ eV applied to N $2p$ on LS [Fe(CNH)$_6$]$^{2+}$ in ONETEP.
\begin{lstlisting}[language=bash]

#--------------------------------------------------------------------------------#
#                             Fe(II)CNH Molecule                                 #
#                           Linear Response J on N                               #
#                                 beta=-0.05                                     #
#--------------------------------------------------------------------------------#

# Parallelization parameters =======================
threads_max                          : 4  # = OMP_NUM_THREADS
threads_num_fftboxes                 : 4  # = OMP_NUM_THREADS
threads_per_fftbox                   : 1  # = 1 (Recommended)
threads_per_cellfft                  : 4  # = threads_max (Recommended, speculated as 14 on Boyle)
threads_num_mkl                      : 1
comms_group_size                     : -1

# Run parameters ===================================
task                                 : SINGLEPOINT
output_detail                        : VERBOSE
xc_functional                        : PBE
PAW                                  : T
check_atoms                          : F
do_properties                        : T
popn_calculate                       : T
turn_off_hartree                     : F
edft                                 : T
edft_smearing_width                  : 0 K

# Cutoff Energy ====================================                                        
cutoff_energy                        : 1088.4563181 eV

# Inner Loop Density Kernel Parameters =============
maxit_lnv                            : 10
minit_lnv                            : 5

# Outer Loop: NGWFs ================================                                                   
ngwf_threshold_orig                  : 2.0e-6
maxit_ngwf_cg                        : 40
delta_e_conv                         : T
elec_energy_tol                      : 2.7211e-8 eV   #2.7211e-8 eV = 1.0d-9 Ha

# Electronic system-specific options ===============
maxit_palser_mano                    : -1
charge                               : 2

# Spin =============================================                                          
spin_polarized                       : T
spin                                 : 0

# Hubbard Parameters ===============================                                              
hubbard_unify_sites                  : F
hubbard_calculating_U                : T
hubbard_ngwf_spin_threshold          : 1.0e-20

# Writing/Reading Variables ========================
write_initial_radial_ngwfs           : F
write_denskern                       : F
write_tightbox_ngwfs                 : F
write_density_plot                   : F
write_ngwf_plot                      : F
write_xyz                            : F
lumo_dens_plot                       : -1
homo_dens_plot                       : -1
lumo_plot                            : -1
homo_plot                            : -1
read_denskern                        : F 
read_tightbox_ngwfs                  : F
print_qc                             : F
write_forces                         : T
cube_format                          : F
grd_format                           : F

# Blocks ===========================================
%block species_ngwf_plot
NU
Fe
C
N
H
%endblock species_ngwf_plot

%block species
bohr
NU N  7  8  12.0
Fe Fe 28 26 14.0
C  C  6  8  12.0
N  N  7  8  12.0
H  H  1  2  10.0
%endblock species

%block species_atomic_set
NU "SOLVE conf=2s2:0.2 2p3:0.2"
Fe "SOLVE conf=3s2:0.2 3p6:0.2 3d6:0.2 4s2:0.2 4p0:0.2 INIT SPIN=0 CHARGE=+2"
C  "SOLVE conf=2s2:0.2 2p2:0.2"
N  "SOLVE conf=2s2:0.2 2p3:0.2"
H  "SOLVE conf=1s1:0.2"
%endblock species_atomic_set

%block species_pot
NU "N.PBE-paw.abinit"
Fe "Fe.PBE-paw.abinit"
C  "C.PBE-paw.abinit"
N  "N.PBE-paw.abinit"
H  "H.PBE-paw.abinit"
%endblock species_pot

%block lattice_cart
ang
20.0 0.00 0.00
0.00 20.0 0.00
0.00 0.00 20.0
%endblock lattice_cart

# Hubbard Parameters (species,L,U,J,projector,alpha,sigma)
%block hubbard
NU 1 0 0 -1.0 0.00 0.1000
Fe 2 0 0 -1.0 0.00 0.00
C  1 0 0 -1.0 0.00 0.00
N  1 0 0 -1.0 0.00 0.00
H  0 0 0 -1.0 0.00 0.00
%endblock hubbard                       

# Atomic Positions
%block positions_abs
ang
NU     9.999999283279960      6.942433250710800      10.00000032390771    #8th
Fe     10.00000016356901      10.00000001883341      10.00000004388317
C      9.999999879327980      11.90392164175866      10.00000018251570
C      8.096476812233410      9.999999468948550      9.999999735475170
C      10.00000038770608      10.00000054897913      8.096476720305730
C      10.00000035596704      10.00000052048448      11.90352333877423
C      11.90352346690779      9.999999374800180      9.999999734105970
C      9.999999746884190      8.096078329933050      10.00000018570493
H      9.999998807597550      5.936632602972260      10.00000033969773
N      6.942816779870070      9.999998827063230      9.999999752303930
H      5.937021325058640      9.999998082666460      9.999999512695780
N      10.00000044427049      10.00000109541506      6.942816688592810
H      10.00000065305725      10.00000172402335      5.937021237120600
N      13.05718350009071      9.999998814131110      9.999999684753950
H      14.06297894433273      9.999998340708830      9.999999417631850
N      10.00000038178598      10.00000113150434      13.05718336221952
H      10.00000054414653      10.00000184628422      14.06297878499432
N      9.999999380427680      13.05756671909915      10.00000035553131
H      9.999998843486900      14.06336736168372      10.00000049978560
%endblock positions_abs


#--------------------------------------------------------------------------------
#-------------------------------- END INPUT FILE --------------------------------
#--------------------------------------------------------------------------------

\end{lstlisting}

\newpage
\subsection{\label{SI_PDOS}PDOS: Hubbard correction to all valence subspaces or Fe $3d$ only}

\begin{figure}[hp]
    \centering
    \includegraphics[width=0.8\linewidth]{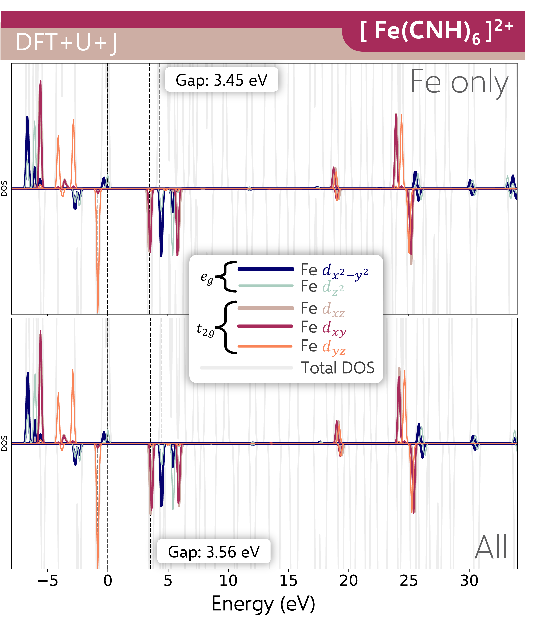} 
    \caption{Projected density of states (PDOS) plots and HOMO-LUMO gaps of iron for HS [Fe(CNH)$_6$]$^{2+}$ using the Himmetoglu DFT+U+J functional equipped with \insitu\ Hubbard parameters applied to Fe only (top) or all valences subspaces in the molecule (All). The e$_g$ orbitals comprise the Fe $d_{x^2-y^2}$ (blue) and $d_{z^2}$ (green) orbitals, and t$_{2g}$ comprise the Fe $d_{xy}$ (pink), $d_{xz}$ (tan), and $d_{yz}$ (orange) orbitals. Total DOS is shown in light gray. Dashed black lines indicate frontier (HOMO or LUMO) orbital energies, and the spin-up and spin-down frontier orbital energies are also indicated by light gray dashed lines.}
    \label{SI_CNH_PDOS}
\end{figure}

\newpage
\subsection{\label{SI_PDOS}Densities: correction to all valence subspaces on Fe $3d$ only}

\begin{figure}[hp]
    \centering
    \includegraphics[width=0.7\linewidth]{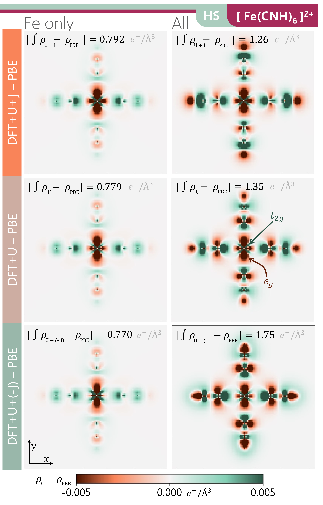} 
    \caption{Cross-section at $z=10$ \AA\ (bisecting the iron atom) of the difference in density (in $e^-/\textrm{\AA}^3$) of Hubbard functional $f$ with respect to PBE for [Fe(CNH)$_6$]$^{2+}$ in the HS state. Left column corresponds to Hubbard corrections applied to the Fe $3d$ orbitals only (Fe only) and right column to Hubbard corrections applied to all valence subspaces (All). Absolute value of the integral of that density difference across the entire $20\textrm{\AA}\times20\textrm{\AA}\times20\textrm{\AA}$ cell is written explicitly.}
    \label{SI_CNH_HS_densities}
\end{figure}

\begin{figure}[hp!]
    \centering
    \includegraphics[width=0.7\linewidth]{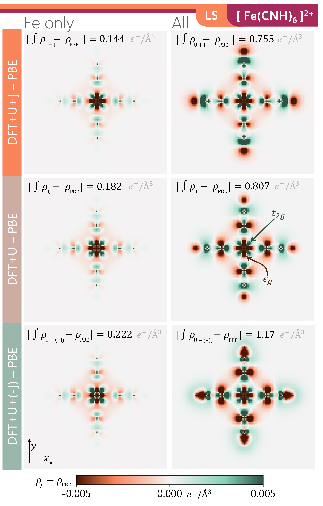} 
    \caption{Cross-section at $z=10$ \AA\ (bisecting the iron atom) of the difference in density (in $e^-/\textrm{\AA}^3$) of Hubbard functional $f$ with respect to PBE for [Fe(CNH)$_6$]$^{2+}$ in the LS state. Left column corresponds to Hubbard corrections applied to the Fe $3d$ orbitals only (Fe only) and right column to Hubbard corrections applied to all valence subspaces (All). Absolute value of the integral of that density difference across the entire $20\textrm{\AA}\times20\textrm{\AA}\times20\textrm{\AA}$ cell is written explicitly.}
    \label{SI_CNH_LS_densities}
\end{figure}

\newpage
\subsection{\label{SI_CommonEnergetics}Energetics for Hubbard functionals with spin-state-specific HPs}

\begin{figure}[ht]
    \centering
    \includegraphics[width=1.0 \linewidth]{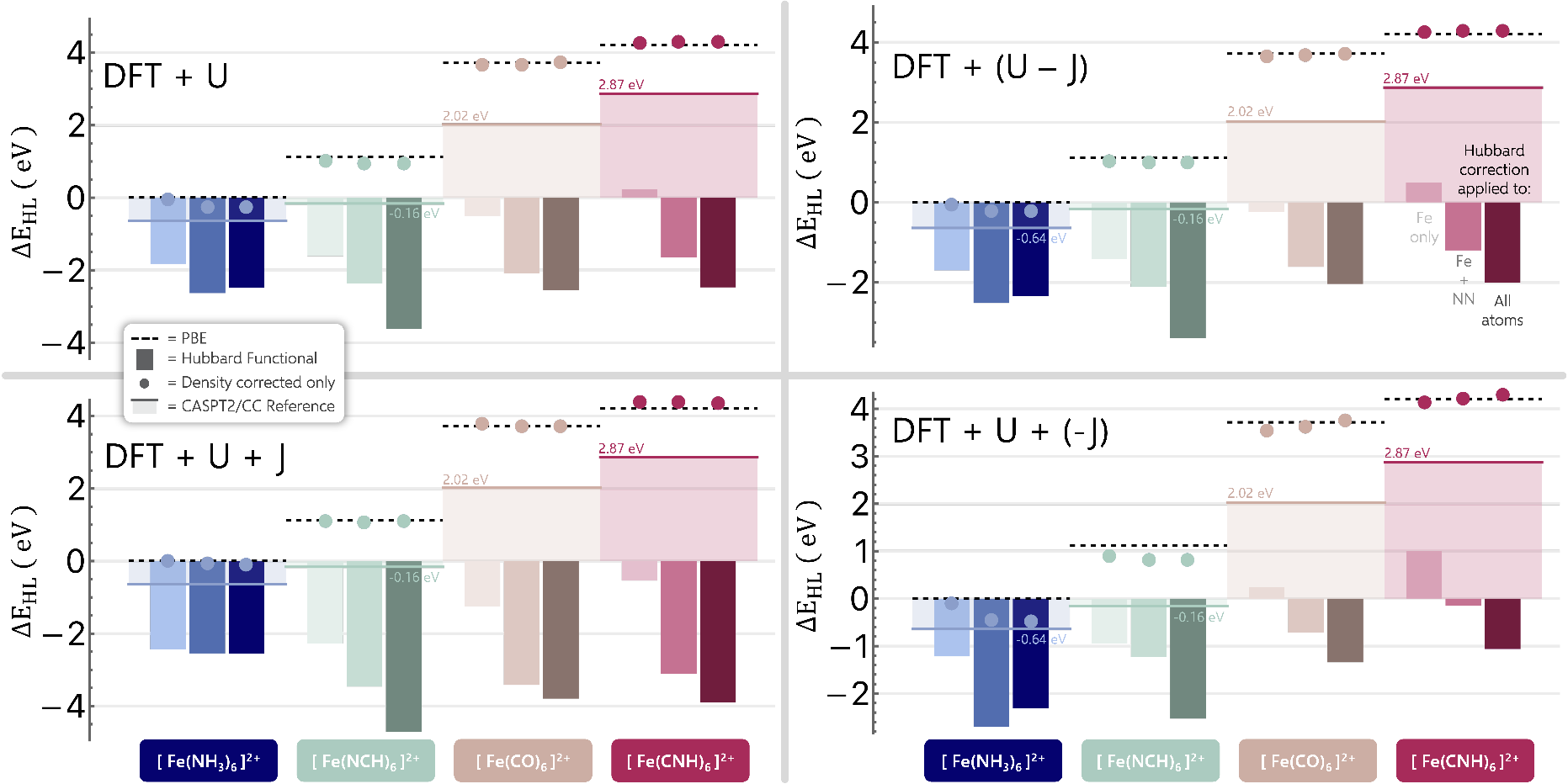} 
    \caption{Adiabatic energy differences \DeltaE\ for all molecules obtained via the denoted Hubbard functional (Dudarev \textrm{DFT+U} upper left; \textrm{DFT+U$-$J} upper right; Himmetoglu \textrm{DFT+U+J} lower left; \textrm{DFT+U+(-J)} lower right), where correction is applied to some combination of subspaces (\textrm{Fe} = \textrm{Fe} $3d$ only, \textrm{Fe+NN} = \textrm{Fe} $3d$ and \textrm{NN} $2p$, \textrm{All} = all valence subspaces in molecule), calculated with the spin-state-specific (\insitu) \textrm{\{U, J\}} parameter pairs. Values are compared to PBE (black dashed lines) and \textrm{CASPT2/CC} reference values (solid color line with shading). \textrm{“Density-corrected”} values (data points) are the PBE@$f$ total energy differences converged with the denoted Hubbard functional $f$, but Hubbard energy corrections are removed non-self-consistently.}
    \label{CommonEnergeticsInSitu}
\end{figure}

\newpage
\subsection{\label{SI_PbeatPBEU}Ranking of MAE density-corrected Hubbard functionals}

\setlength\arrayrulewidth{1pt}
\setlength{\extrarowheight}{0pt}
\begin{table}[h!]
\footnotesize
\begin{tabular}{r|cc|cccc|c|c}
\multicolumn{1}{l}{} & \multicolumn{2}{c}{HP specs} & \multicolumn{4}{c}{$\Delta E_\tr{HL}$ (eV)} & \multicolumn{1}{c}{Error (eV)} & \multicolumn{1}{l}{} \\ \cmidrule(lr){2-3}\cmidrule(lr){4-7}\cmidrule(lr){8-8}
Functional & Atoms & \{U, J\} & \cellcolor[HTML]{04017A}{\color[HTML]{FFFFFF} NH3} & \cellcolor[HTML]{A9CDC0}{\color[HTML]{FFFFFF} NCH} & \cellcolor[HTML]{CCADA4}{\color[HTML]{FFFFFF} CO} & \cellcolor[HTML]{A82B5A}{\color[HTML]{FFFFFF} CNH} & & $\overline{\textrm{MAE}}$ \\ \hline
\rowcolor[HTML]{EEF2F4}[\overhang]
CASPT2/CC & -- & -- & -0.64 & -0.16 & 2.02 & 2.87 & \cellcolor{white}{} & 0.00 \\ \hline
\multicolumn{1}{|r|}{PBE@U + (-J)} & {\color[HTML]{808080} Fe+NN} & \cellcolor[HTML]{EDEDEF}\insitu & -0.45 & 0.81 & 3.61 & 4.20 &  & \multicolumn{1}{c|}{1.02} \\
PBE@U + (-J) & All & \cellcolor[HTML]{EDEDEF}\insitu & -0.47 & 0.81 & 3.75 & 4.31 &  & 1.08 \\
PBE@U + (-J) & {\color[HTML]{BFBFBF} Fe} & \cellcolor[HTML]{EDEDEF}\insitu & -0.11 & 0.91 & 3.54 & 4.13 &  & 1.09 \\
PBE@U & {\color[HTML]{808080} Fe+NN} & \cellcolor[HTML]{EDEDEF}\insitu & -0.26 & 0.96 & 3.68 & 4.31 &  & 1.15 \\
PBE@U-J & {\color[HTML]{808080} Fe+NN} & \cellcolor[HTML]{EDEDEF}\insitu & -0.20 & 0.99 & 3.67 & 4.29 &  & 1.17 \\
PBE@U & All & \cellcolor[HTML]{EDEDEF}\insitu & -0.26 & 0.96 & 3.73 & 4.32 &  & 1.17 \\
PBE@U-J & All & \cellcolor[HTML]{EDEDEF}\insitu & -0.21 & 1.00 & 3.70 & 4.28 &  & 1.17 \\
PBE@U + (-J) & {\color[HTML]{808080} Fe+NN} & HS & -0.30 & 0.99 & 3.77 & 4.33 &  & 1.18 \\
PBE@U + (-J) & {\color[HTML]{BFBFBF} Fe} & HS & -0.07 & 0.99 & 3.70 & 4.24 &  & 1.19 \\
PBE@U-J & {\color[HTML]{BFBFBF} Fe} & \cellcolor[HTML]{EDEDEF}\insitu & -0.05 & 1.02 & 3.66 & 4.26 &  & 1.20 \\
PBE@U & {\color[HTML]{BFBFBF} Fe} & \cellcolor[HTML]{EDEDEF}\insitu & -0.05 & 1.00 & 3.67 & 4.29 &  & 1.20 \\
PBE & -- & -- & 0.01 & 1.09 & 3.72 & 4.21 &  & 1.23 \\
PBE@U + (-J) & All & HS & -0.33 & 0.99 & 3.92 & 4.48 &  & 1.24 \\
PBE@U+J & All & \cellcolor[HTML]{EDEDEF}\insitu & -0.10 & 1.11 & 3.72 & 4.36 &  & 1.25 \\
PBE@U+J & {\color[HTML]{808080} Fe+NN} & \cellcolor[HTML]{EDEDEF}\insitu & -0.07 & 1.08 & 3.73 & 4.38 &  & 1.26 \\
PBE@U & {\color[HTML]{808080} Fe+NN} & HS & -0.12 & 1.09 & 3.82 & 4.40 &  & 1.28 \\
PBE@U-J & {\color[HTML]{808080} Fe+NN} & HS & -0.07 & 1.10 & 3.80 & 4.37 &  & 1.28 \\
PBE@U-J & {\color[HTML]{BFBFBF} Fe} & HS & -0.01 & 1.10 & 3.78 & 4.34 &  & 1.28 \\
PBE@U & {\color[HTML]{BFBFBF} Fe} & HS & -0.01 & 1.09 & 3.80 & 4.36 &  & 1.29 \\
PBE@U-J & All & HS & -0.07 & 1.10 & 3.84 & 4.40 &  & 1.30 \\
PBE@U+J & {\color[HTML]{BFBFBF} Fe} & \cellcolor[HTML]{EDEDEF}\insitu & 0.01 & 1.10 & 3.78 & 4.4 &  & 1.30 \\
PBE@U & All & HS & -0.12 & 1.09 & 3.89 & 4.46 &  & 1.31 \\
PBE@U+J & {\color[HTML]{808080} Fe+NN} & HS & 0.03 & 1.18 & 3.87 & 4.46 &  & 1.36 \\
PBE@U+J & All & HS & 0.03 & 1.18 & 3.89 & 4.47 &  & 1.37 \\
\multicolumn{1}{|r|}{PBE@U+J} & {\color[HTML]{BFBFBF} Fe} & HS & 0.05 & 1.18 & 3.89 & 4.48 & \multirow{-27}{*}{\hspace{-0.0cm}\includegraphics[width=0.213\linewidth,trim={0.23cm 0cm 0.25cm 0.105cm},clip]{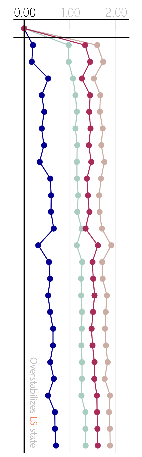}} & \multicolumn{1}{c|}{1.38} \\ \hline
\end{tabular}
\caption{Ranking of common \textrm{PBE@$f$} density-corrected Hubbard functionals in terms of mean average errors (\textrm{MAE}) for adiabatic energy differences $\Delta E_\tr{HL}$, averaged across all complexes (MAE), with respect to CASPT2/CC reference across all molecules (first row).\cite{Mariano2020,Mariano2021} Histogram shows signed error with respect to CASPT2/CC values. \textrm{\{U, J\}} column delineates use of either HS state HP pairs or \insitu\ pairs for each respective spin state. \textrm{“Atoms”} column indicates to which subspaces corrections were applied (\textrm{Fe} = \textrm{Fe} $3d$ only, \textrm{Fe+NN} = \textrm{Fe} $3d$ and \textrm{NN} $2p$, \textrm{All} = all valence subspaces in molecule). Bracketed rows depict the range of average MAE obtained through the density-corrected \textrm{PBE@$f$} functionals, with the outermost (most and least accurate of the \textrm{PBE@$f$} functionals) given rows themselves.}
\label{PBE@PBE+U}
\end{table}

\newpage
\newpage
\bibliography{SI}